\documentclass[aps,prx,floatfix,amsmath,amssymb,preprint,eqsecnum,nofootinbib,superscriptaddress]{revtex4}

\usepackage{graphicx, bm, amsmath, amsfonts,amssymb,amsthm}

\usepackage{float}
\usepackage{caption}
\usepackage{color}
\usepackage{ulem}

\usepackage{array,amsmath,amsthm,amssymb,amsfonts,graphicx,setspace,color,verbatim,wrapfig,hyperref,mathtools,stackengine}

\title{}
\date{} 

\newcommand{\nn}[2]{\langle #1 #2 \rangle}

\renewcommand{\d}{\partial}

\renewcommand{\nn}{\nonumber}

\newcommand{\beq}{\begin{equation}}
\newcommand{\eeq}{\end{equation}}
\newcommand{\ba}{\begin{array}{ccc}}
\newcommand{\ea}{\end{array}}
\newcommand{\bea}{\begin{eqnarray}}
\newcommand{\eea}{\end{eqnarray}}

\newcommand{\cG}{{\cal{G}}}

\newtheorem{prop}{Proposition}

\graphicspath{{figures/}}

\begin{document}

\title{1d lattice models for the boundary of 2d ``Majorana" fermion SPTs: Kramers-Wannier duality as an exact $Z_2$ symmetry.}

\author{Robert A. Jones}
\affiliation{Department of Physics, Massachusetts Institute of Technology, Cambridge, MA  02139, USA}

\author{Max A. Metlitski}
\affiliation{Department of Physics, Massachusetts Institute of Technology, Cambridge, MA  02139, USA}

\date{\today}

\begin{abstract}
It is well known that symmetry protected topological (SPT) phases host non-trivial boundaries that cannot be mimicked in a lower-dimensional system with a conventional realization of symmetry. However, for SPT phases of bosons (fermions) within the cohomology (supercohomology) classification the boundary can be recreated without the bulk at the cost of a non-onsite symmetry action. This raises the question: can one also mimic the   boundaries of SPT phases which lie {\it outside} the (super)cohomology classification? In this paper, we study this question in the context of 2+1D fermion SPTs. We focus on the root SPT phase for the symmetry group $G  =Z_2 \times Z^f_2$. Starting with an exactly solvable model for the bulk of this phase constructed by Tarantino and Fidkowski, we derive an effective 1d lattice model for the boundary. Crucially, the Hilbert space of this 1d model does not have a local tensor product structure, but rather is obtained by placing a local constraint on a local tensor product Hilbert space. We derive the action of the $Z_2$ symmetry on this Hilbert space and find a simple 3-site Hamiltonian that respects this symmetry. We study this Hamiltonian numerically using exact diagonalization and DMRG and find strong evidence that it realizes an Ising CFT where the $Z_2$ symmetry acts as the Kramers-Wannier duality; this is the expected stable gapless boundary state of the present SPT. A simple modification of our construction realizes the boundary of the 2+1D topological superconductor protected by time-reversal symmetry ${\cal T}$ with ${\cal T}^2 = (-1)^{\cal F}$. 
\noindent

\end{abstract}

\maketitle
\newpage

\section{Introduction}
Symmetry protected topological (SPT) phases have attracted a lot of attention in recent years.\cite{Chen2012, SenthilSPTRev}  A key property of SPT phases is the presence of non-trivial boundary states protected by a symmetry group $G$: as long as the symmetry is not explicitly broken, a  gapped symmetric boundary state with no intrinsic topological order is prohibited. Furthermore, the boundary of an SPT phase is anomalous: it cannot be mimicked in a strictly lower-dimensional system with a conventional Hilbert space and realization of symmetry. Here by a conventional Hilbert space we mean a Hilbert space $V$ with a local tensor product structure: i.e. $V = \otimes_i V_i$, where $i$ labels the ``sites" of the lattice, and $V_i$ - the local site Hilbert space. Likewise, by a  conventional realization of symmetry we mean that the symmetry is ``onsite,"\cite{Chen2013} i.e. for each group element $g \in G$ its action factorizes into a product over sites 
\beq U(g) = \otimes_i U_i(g) \label{onsite} \eeq
 with $U_i(g)$ obeying the group law.\footnote{Unless specifically noted, we don't consider space-group symmetries in this paper.} However, it is known that the boundary of {\it some} SPT phases can be recreated without the bulk, provided that one relaxes the assumption of onsite symmetry action (\ref{onsite}) while keeping the assumption of the local tensor product Hilbert space. This is true for boson SPT phases in the cohomology classification\cite{Chen2013, CX, CZX, ElseNayak}, and believed to be true for fermion SPT phases in the supercohomology classification.\cite{supercohomology, ElseNayak, TylerLukasz, NatAshvin} In these cases, the non-onsite symmetry is a finite depth local unitary, which, however, cannot be factorized as (\ref{onsite}). Furthermore, given a non-onsite action of the symmetry in the effective boundary model one can extract the algebraic data that defines the corresponding bulk SPT phase.\cite{ElseNayak}\footnote{For bulk spatial dimension $d \ge 3$ this is subject to assuming a certain ansatz for the form of the non-onsite symmetry action.\cite{ElseNayak}}
 
It is known that there exist SPT phases which are not part of the  (super)cohomology classification.
For bosons, the first such phase appears in three spatial dimensions\footnote{Unless otherwise noted, dimensions stand for spatial dimensions.}; its protecting symmetry is time-reversal.\cite{Vishwanath2013,Burnell2013} For fermions, such phases exist already in two dimensions with unitary symmetry:  the simplest example, which will be the main subject of this paper, is provided by the symmetry group $G = Z_2 \times Z^f_2$.\footnote{$Z^f_2$ is the  fermion parity symmetry.}\cite{supercohomology,GK,Bhardwaj} One can ask whether the boundaries of such beyond (super)cohomology phases can also be recreated without the bulk and, if so, what assumptions about the form of the effective boundary Hilbert space and symmetry action need to be sacrificed. 
In this paper, we will construct a lattice model for the 1d edge of the beyond supercohomology 2d fermion SPT phase with $G = Z_2 \times Z^f_2$. 
Unlike for (super)cohomology SPTs, our effective edge model lives in a Hilbert space which does not have a local tensor product structure, but rather is obtained from a local tensor product Hilbert space by placing a local constraint. Our construction trivially generalizes to all 2d fermion SPTs with symmetry group $G = G_b \times Z^f_2$. It also extends to the 2d topological superconductor with time-reversal symmetry ${\cal T}$ and ${\cal T}^2 = (-1)^F$. 

Let us review the properties of 2d fermion SPTs  with $Z_2 \times Z^f_2$ symmetry. In the absence of interactions, such fermion phases are classified by an integer $n \in Z$; interactions reduce the classification to $n \in Z_8$.\cite{Gu_Levin} The generator of the classification $n =1$ can be obtained by stacking a $p+ip$ and a $p-ip$ superconductor, where only the fermions in the $p+ip$ layer are charged under $Z_2$. Correspondingly, the 1d edge of this  phase hosts a pair of counter-propagating Majorana ($c  =1/2$) edge modes
\beq H^{edge} = -i( \chi_R  \d_x\chi_R - \chi_L \d_x \chi_L) \label{HIsing}\eeq
 where only the right-mover $\chi_R$ is charged under the $Z_2$ symmetry,
 \beq Z_2: \chi_R \to -\chi_R, \quad \chi_L \to \chi_L \label{Z2intro} \eeq
 From the Ising CFT standpoint, this $Z_2$ is the Kramers-Wannier (KW) self-duality symmetry; the mass term 
 \beq L_m = i m \chi_R \chi_L \label{Lm} \eeq
  which drives the phase transition in the Ising model, is odd under this symmetry, thus, the edge is automatically tuned to the self-dual critical point. Thus, to mimic the boundary in 1d,  we need an Ising model on the lattice with an exact $Z_2$ self-duality symmetry.
  
One may think that the standard KW duality of the transverse field Ising model (TFIM) does the job. However, this is not the case. The generator of the KW duality squares to a {\it translation} in the TFIM and thus, generates a $Z$ symmetry rather than a $Z_2$ symmetry. Indeed, consider the Hamiltonian of the TFIM:
\beq H_{TFIM} = -J \sum_{i} \sigma^x_i \sigma^x_{i+1} + h \sum_i \sigma^z_i \label{HTFIM}\eeq
The standard KW duality maps $H_{TFIM}$ to a Hamiltonian on the dual lattice via the transformation:
\beq \mu^z_{i + 1/2}  = -\sigma^x_i \sigma^x_{i+1}, \quad \mu^x_{i-1/2} \mu^x_{i+1/2} = -\sigma^z_i\eeq
which interchanges $J$ and $h$. The self-dual point is $J  = h$. If we want to treat the KW duality at $J = h$ as a symmetry, we need a way to identify the dual lattice and the direct lattice. For instance, we can shift the dual lattice by  half a unit cell and let the duality transformation $U_{KW}$ act as
\beq U_{KW} \sigma^z_i U^{\dagger}_{KW} = -\sigma^x_{i} \sigma^x_{i+1}, \quad
U_{KW} \sigma^x_i \sigma^x_{i+1} U^{\dagger}_{KW}=-\sigma^z_{i+1}\eeq
Then $U^2_{KW}$ is just the translation by one lattice site. It is also useful to think about this in the fermionic  language (via the Jordan-Wigner transformation):
\beq H_{Maj} = i h \sum_i \gamma_i \bar{\gamma}_i + i J \sum_{i} \bar{\gamma}_i \gamma_{i+1} \label{Hmaj}\eeq
where $\gamma_i$ and $\bar{\gamma}_i$ are Majorana operators sitting on the sites of the direct lattice. We have
\beq U_{KW} \gamma_i U^{\dagger}_{KW} = \bar{\gamma}_i, \quad U_{KW} \bar{\gamma}_i U^{\dagger}_{KW} = \gamma_{i+1} \label{UKWg}\eeq
If we define $\gamma_{i+1/2} = \bar{\gamma}_i$, then $U_{KW}$ is just  the translation by half a unit cell, so $U^2_{KW}$ is a translation by one unit cell.

When we diagonalize the Majorana chain (\ref{Hmaj}) at $J = h$, we obtain the effective low energy theory (\ref{HIsing}). The right-mover $\chi_R$ is localized at momentum $\pi$ and the left-mover $\chi_L$ at momentum $0$,\footnote{Here we give the momentum with respect to the translation by half a unit cell.} so at low energy $U_{KW}$ acts precisely via Eq.~(\ref{Z2intro}) - i.e. like an internal $Z_2$ symmetry,\footnote{This has been utilized for numerical simulations of SPT edge in Ref.~\onlinecite{GroverSheng}.} however, this is not true at the lattice scale. 


  
\begin{figure}[t]
\includegraphics[width = 0.5\linewidth]{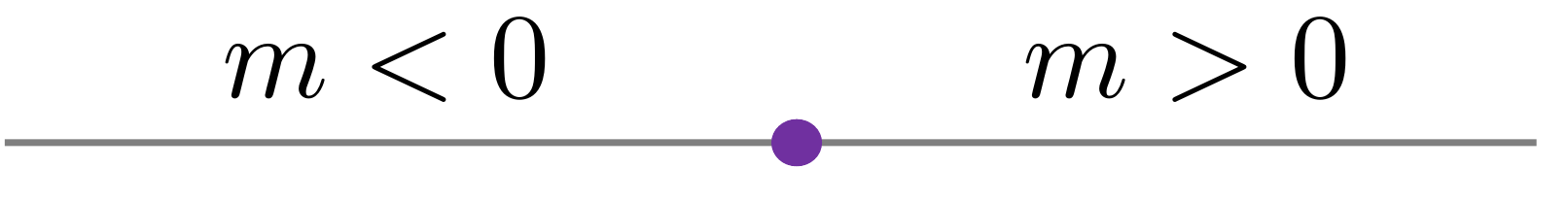}
\caption{Boundary of the $n  =1$ $Z_2\times Z^f_2$ SPT: a domain wall between two opposite domains of the $Z_2$ symmetry carries a Majorana mode.}
\label{fig:DW}
\end{figure}

 We thus ask: can one mimick the edge of the  $n =1$ $Z_2\times Z^f_2$ SPT keeping the symmetry as an {\it internal} $Z_2$ symmetry. One can argue that this cannot be achieved with a (fermionic) local tensor product Hilbert space and a $Z_2$ symmetry that acts as a locality preserving unitary, i.e. sacrificing just the assumption of onsite symmetry (\ref{onsite}) is not sufficient.\cite{ElseNayak} By a locality preserving unitary we mean a unitary that maps local operators to local operators.  Indeed, imagine breaking the $Z_2$ symmetry with the perturbation (\ref{Lm}): the edge becomes gapped. A domain wall between  regions with $m > 0$ and $m < 0$ traps a Majorana zero mode, see Fig.~\ref{fig:DW}. From this one concludes that edge phases with $m < 0$ and $m > 0$ effectively differ by a Kitaev chain. Yet, they are mapped into each other by the $Z_2$ symmetry $U$, which must then effectively paint a Kitaev chain on the boundary. But a Kitaev chain is a non-trivial 1d phase of fermions, so it cannot be created by a finite depth local unitary. It can, however, be created by a non-finite depth local unitary: the half-translation operator implementing (\ref{UKWg}) is one example. In fact, it was argued in Ref.~\onlinecite{FloquetF} that in 1d fermion systems all locality preserving unitaries modulo finite depth local unitaries are classified by an index $\nu_f =  \zeta \log \sqrt{2} + \log \frac{p}{q}$, where $\zeta = 0, 1$ and $p$, $q$ are positive integers.\footnote{See also the preceding work in Refs.~\onlinecite{FloquetR, FloquetC, Automata}.}  For instance, the half-translation (\ref{UKWg}) has $\nu_f = \log \sqrt{2}$. For a locality preserving unitary  which is not finite depth, $\nu_f \neq 0$. Furthermore, for two locality preserving unitaries $U_1$, $U_2$, $\nu_f(U_1 U_2) = \nu_f(U_1) + \nu_f(U_2)$. Thus, the locality preserving unitary $U$ implementing our $Z_2$ symmetry must have $\nu_f \neq 0$ and so $\nu_f(U^2)  \neq 0$, i.e. $U^2  \neq 1$, and, in fact,  $U^p \neq 1$ for any finite power $p$.


 Thus, the effective 1d lattice model for the boundary of the $n = 1$ phase must be qualitatively different from that of supercohomology phases with even $n$ discussed in Refs.~\onlinecite{TylerLukasz,NatAshvin}.  In fact, the above discussion hints at the form the boundary model should take. We can imagine starting with a $Z_2$ symmetry broken state on the boundary and introducing domains of positive and negative mass $m$.  If we want the model to be capable of describing  symmetry-preserving states, the domain walls should be mobile. It should also be possible to create and destroy them in pairs. Since each domain wall traps a Majorana mode, the model must describe a Majorana liquid.
 
In this paper, we derive a 1d lattice model providing a ``bosonized" description of such a Majorana liquid, which is somewhat akin to the Jordan-Wigner bosonization of the Majorana chain (\ref{Hmaj}). Our starting point is the model for the $n = 1$ $Z_2 \times Z^f_2$ SPT introduced by Tarantino and Fidkowski (TF) in Ref.~\onlinecite{FidkowskiSpin}.\footnote{See also Ref.~\onlinecite{Bhardwaj} and the closely related model in Ref.~\onlinecite{Bela}.} Unlike the free fermion description of this phase discussed above, the TF model is a (strongly interacting) commuting projector Hamiltonian. While the ground state of the TF model is unique on a closed manifold, it is highly degenerate on a manifold with a boundary: the degeneracy grows exponentially with the boundary length. As a result, the effective boundary Hilbert space separates cleanly from the bulk (unlike for the free-fermion model where the chiral edges necessarily connect the bulk bands). We find the action of the $Z_2$ symmetry in the boundary Hilbert space and present a simple 3-site Hamiltonian consistent with this symmetry. Our numerical exact diagonalization and DMRG studies strongly suggest that this boundary Hamiltonian flows to an Ising CFT, where the $Z_2$ symmetry acts via Eq.~(\ref{Z2intro}).

This paper is organized as follows. In Section \ref{sec:model} we define our 1d edge model and study it numerically. In section \ref{sec:der} we give a derivation of this effective edge model starting from the TF model for the bulk $Z_2 \times Z^f_2$ SPT. Concluding remarks are given section \ref{sec:disc}. 
 
 
\section{1d model}
\label{sec:model}
We begin by introducing the boundary lattice model of the $n = 1$ $Z_2 \times Z^f_2$ SPT; we postpone its derivation to section \ref{sec:der} where the intuitive justifications made in this section will be made precise. 
\subsection{Hilbert space}
\label{sec:Hilbert}

\begin{figure}[t]
\includegraphics[width = \linewidth]{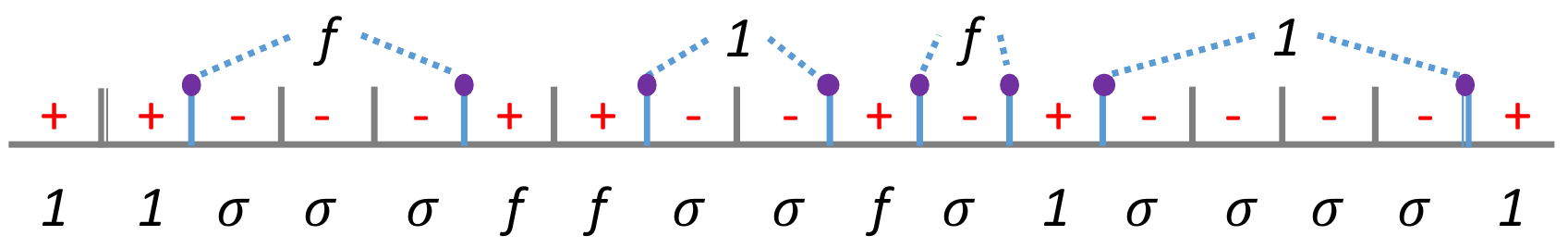}
\caption{An example of an admissible state on a segment (segment boundaries are shown with double lines). Each link is labelled as $1$, $\sigma$ or $f$. The leftmost link corresponds to $i = 0$ and the rightmost link to $i = N_{seg}+1$. Domain walls are marked in solid blue: each domain wall supports a Majorana fermion (purple circles). The fusion product of boundary Majoranas on each $``-"$ domain (top line) is determined by the product of the $1, f$ labels on the $``+"$ domains to the immediate left and right. }
\label{fig:Segment}
\end{figure}

We first describe the boundary Hilbert space.  We work on a circle and consider a 1d array with $N$ sites arranged periodically. We begin with a Hilbert space, which is a tensor product over the bonds of this array. The Hilbert space on each bond is spanned by  three states labelled  $\{1, \sigma, f\}$. The labels are suggestively chosen to be the same as  the anyon types in the Ising  theory. We will  also sometimes denote these as $\{1, \sigma, f\} \sim \{0, \frac12, 1\}$. We then impose a  constraint that $1$ and $f$ cannot sit on adjacent bonds. Physically, this Hilbert space has the following interpretation. We think of each bond as a microscopic domain of the $Z_2$ symmetry: it carries an Ising spin of $``+"$ or $``-"$. 
We know that every domain wall between a $``+"$ and a $``-"$ domain carries a Majorana mode: if we have $N_{dw}$ domain walls, the Majorana modes span a subspace of dimension $2^{N_{dw}/2}$. To describe the state, it is not sufficient to just specify all the Ising spins, we  must also specify the state in the Majorana subspace. It is simpler to do this when our system is a line segment rather than a circle (we will return to discuss the case of the circle shortly). We can embed a segment in a circle, and choose  the complement of the segment to have all Ising spins frozen at $``+"$. This corresponds to boundary conditions on the segment which break the $Z_2$ symmetry. We let the bonds of the segment be numbered as $i =1 \ldots N_{seg}$. Now, for a fixed configuration of Ising spins on the segment, we label each bond by the fusion product of all the Majorana's sitting to the left of the bond. The $``-"$ bonds have an odd number of Majoranas to the left of them, so they necessarily carry the label $\sigma$. The $``+"$ bonds have an even number of Majoranas to the left of them, so they carry labels $1$ or $f$. Thus, we don't have to separately note the Ising spin of the bond - it can be read off from the $1, \sigma, f$ label: ${1, f} \to ``+"$, $\sigma \to ``-"$. The $i  =0$ bond (just to the left of the segment) by convention is labelled as $1$. The $i =N_{seg} + 1$ bond (just to the right of the segment) can be $1$ or $f$ depending on whether the total fermion parity of the state is even or odd. This labeling is essentially a fusion tree of the Majoranas. We notice that with this convention, we can never have a $1$ adjacent to an $f$, as claimed. All other sequences of $1, \sigma, f$, are  allowed. We also point out that this labeling convention essentially corresponds to using a basis for the Majorana subspace obtained by grouping the boundary Majoranas of each $``-"$ domain into a complex fermion. The occupation number of this complex fermion is given by the mod 2 sum of  labels $a_i \in \{0, 1\}$ on the two adjacent $``+"$ domains, see Fig.~\ref{fig:Segment}.

The utility of this labeling convention is that any physical local bosonic operator $b(x)$ will act locally in the effective Hilbert space constructed. Indeed, let's pick a point $y$ on our segment which is far from point $x$. If $0 < y < x$, then $b(x)$ cannot change the topological charge to the left of $y$ for trivial reasons. If $x < y < N_{seg}$, $b(x)$ cannot change the topological charge to the left of $y$ since $b$ is a bosonic operator. 

Let us discuss how the Hilbert space dimension grows with the length of the segment $N_{seg}$. We have
\bea {\rm{dim}} {\cal V}_1(N_{seg}) &=& \frac{1}{4}((1+\sqrt{2})^{N_{seg} + 1} + (1-\sqrt{2})^{N_{seg}+1}+2)\nn\\
{\rm{dim}}{\cal V}_f(N_{seg})&=& \frac{1}{4}((1+\sqrt{2})^{N_{seg} + 1} + (1-\sqrt{2})^{N_{seg}+1}-2) \label{Vdim}\eea
where ${\cal V}_1$ and ${\cal V}_f$ denote the Hilbert spaces of a segment with total fermion parity even and odd respectively. We see that the Hilbert space dimension grows as $(1+\sqrt{2})^{N_{seg}}$ - a stark signature of the fact that the Hilbert space is constrained and does not have a local tensor product structure.

{\it Circle.} Now let us return to the case of periodic boundary conditions. We still label the $``-"$ domains as $\sigma$ and $``+"$ domains as $1$ or $f$. We  still use a basis in the Majorana subspace where the Majoranas on $``-"$ domains are grouped into complex fermions, and let the occupation number of this complex fermion be given by the difference in labels on the two adjacent $``+"$ domains (Fig.~\ref{fig:U11}, top). This, however, leads to two difficulties. First, let us define an operator $S_i$ acting on bond $i$ via
\beq S_i|a_i\rangle = |\bar{a}_i\rangle \equiv  |1-a_i\rangle, \quad S = \prod_i S_i \label{S}\eeq
i.e. $\bar{1} = f$, $\bar{f} = 1$ and $\bar{\sigma} = \sigma$.  Then acting on a state with $S$ (i.e. uniformly interchanging $1$ and $f$, while keeping $\sigma$ unchanged)
results in the same set of occupation numbers for the complex fermions. Second, this labeling only works if the total fermion parity on the circle is even. It turns out that these two difficulties have a unified resolution. 

\begin{figure}[t]
\includegraphics{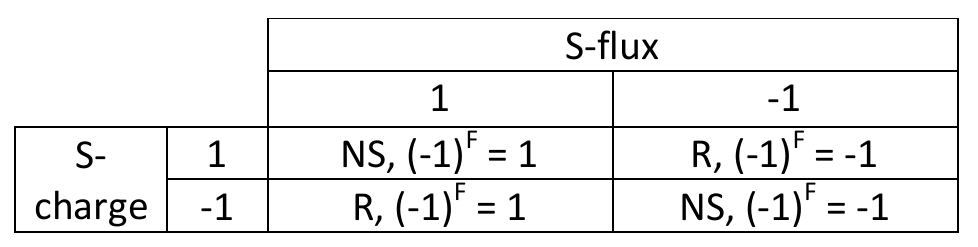}
\caption{Sectors of the model on the circle. The fermion parity symmetry of the microscopic fermion model is related to the $S$ symmetry of our 1d model. NS and R stand for Neveu-Schwarz and Ramond spin-structures of the fermion model. }
\label{fig:NSR}
\end{figure}

Recall that for fermions on a circle we can have two boundary conditions (``spin-structures") which differ by threading fermion parity flux through the circle. In the context of the Ising field theory (\ref{HIsing}) the anti-periodic boundary condition $\chi_{R/L}(x + L) = -\chi_{R/L}(x)$ is known as Neveu-Schwarz (NS) and the periodic boundary condition $\chi_{R/L}(x+L) = \chi_{R/L}(x)$ is known as Ramond (R). In each sector (NS or R) the total fermion parity $(-1)^{\cal F}$ can be either even or odd. Thus, we have four sectors in total. In our effective ``bosonized" description these sectors appear as follows. 
We require bosonic operators of the microscopic fermionic theory to commute with symmetry $S$ in Eq.~(\ref{S}).  $S$ is an ``on-site" symmetry (at least in the local tensor product Hilbert space we start with before placing constraints). We can, thus, put a flux of $S$ through the circle. 
In the ``bosonized" description we have two sectors: sector $1$ with no $S$ flux and sector $2$ with $S$ flux around the circle. Likewise, in each sector we can have $S$ charge  $+1$ or $-1$, see Fig.~\ref{fig:NSR}. The state in sector $1$ with $S  =+1$ corresponds to the NS spin-structure with $(-1)^{\cal F}  =1$. The state in the sector $2$ with $S =-1$ corresponds to the NS spin-structure with $(-1)^{\cal F}  =-1$. The remaining two sectors (sector $1$ with $S  =-1$ and sector $2$ with $S  =1$) correspond to R spin-structure and have opposite fermion parity.  We note that absolute fermion parity in the R sector is ill-defined. Indeed, in the TF model, when the bulk is a disc, the boundary will  be in the NS sector. To access the R sector, we must take a cylinder topology for the bulk and thread fermion parity flux through the hole of the cylinder. The total fermion parity of the two boundaries of the cylinder is defined, but the individual absolute fermion parity of each boundary is a matter of convention. Here we take the convention that sector $1$, to which the state with all spins $``+"$ belongs, has $(-1)^{\cal F} = 1$, while sector $2$, to which the state with all spins $``-"$ belongs, has $(-1)^{\cal F} = -1$.

We note that if we take the perspective of the bosonic Ising model (rather than the fermionic theory) then $S$ is just the Ising symmetry (not to be confused with the self-duality symmetry  (\ref{Z2intro}), which is our main focus here). In fact, the correspondence in Fig.~\ref{fig:NSR} between $S$ charge/flux and the fermion parity charge/spin-structure is identical to that in standard 1d Jordan-Wigner bosonization that maps the Majorana chain (\ref{Hmaj}) to  the TFIM (\ref{HTFIM}). The symmetry $S$ in the TFIM (\ref{HTFIM}) is just the spin-flip symmetry, $S_{TFIM} = \prod_i (-\sigma^z_i)$.


There is a minor subtlety in how to introduce the flux of the $S$ symmetry: this flux affects not only the Hamiltonian, but also the constraint on the  Hilbert space.  Let us place the branch-cut associated with the flux between bonds $i = N$ and $i = 1$. Then we don't allow the sequence  $1, 1$ and $f, f$ on these two bonds; all other bonds have the previous constraint: no $1$ adjacent to an $f$. Below, we will use an equivalent, but more convenient way,  to introduce the $S$-flux: one can work on the double-cover of the circle, i.e. a circle of length $2 N$, where $a_{i + N} = \bar{a}_i$ (Fig.~\ref{fig:U22}, top).  We may schematically label a state on this twisted double-cover as $|a \bar{a}\rangle$, where $a$ now labels a string of length $N$. In the sector with no $S$-twist, we may likewise utilize a double-cover with $a_{i+N} = a_i$ and label a state by $|a a\rangle$ (Fig.~\ref{fig:U21}, top). In this double-cover notation all bonds satisfy no $1$ adjacent to an $f$ rule; the bonds $N$ and $1$ receive no special treatment.

Let $O_i$ be a bosonic operator localized near site $i$. As we already noted, this implies that $O_i$ commutes with $S$. As an example, let's consider an $O_i$ that acts on three sites $i-1$, $i$, $i+1$ (terms in the model Hamiltonian we consider below have this property). Let $O_i$ act on an infinite line via
\beq O_i |\ldots a_{i-1} a_i a_{i+1} \ldots\rangle = \sum_{b_{i-1}, b_{i}, b_{i+1}} O_i(b_{i-1}, b_i, b_{i+1}; a_{i-1}, a_i, a_{i+1}) |\ldots b_{i-1} b_i b_{i+1} \ldots\rangle \nn\eeq
then in the twisted sector in the double-cover notation,
\bea &&O_i |\ldots a_{i-1} a_i a_{i+1} \ldots \bar{a}_{N+i-1} \bar{a}_{N+i} \bar{a}_{N+i+1}\rangle \nn\\
&=& \sum_{b_{i-1}, b_{i}, b_{i+1}} O_i(b_{i-1}, b_i, b_{i+1};a_{i-1}, a_i, a_{i+1}) |\ldots b_{i-1} b_i b_{i+1} \ldots \bar{b}_{N+i-1} \bar{b}_{N+i} \bar{b}_{N+i+1}\ldots\rangle \nn\eea
This corresponds precisely to putting a flux of $S$-around the circle. It is well-defined since $O_i$ on an infinite line is assumed to be $S$-invariant.



\subsection{Symmetry action}
\label{sec:sym}


\begin{figure}[t]
\includegraphics[width = \linewidth]{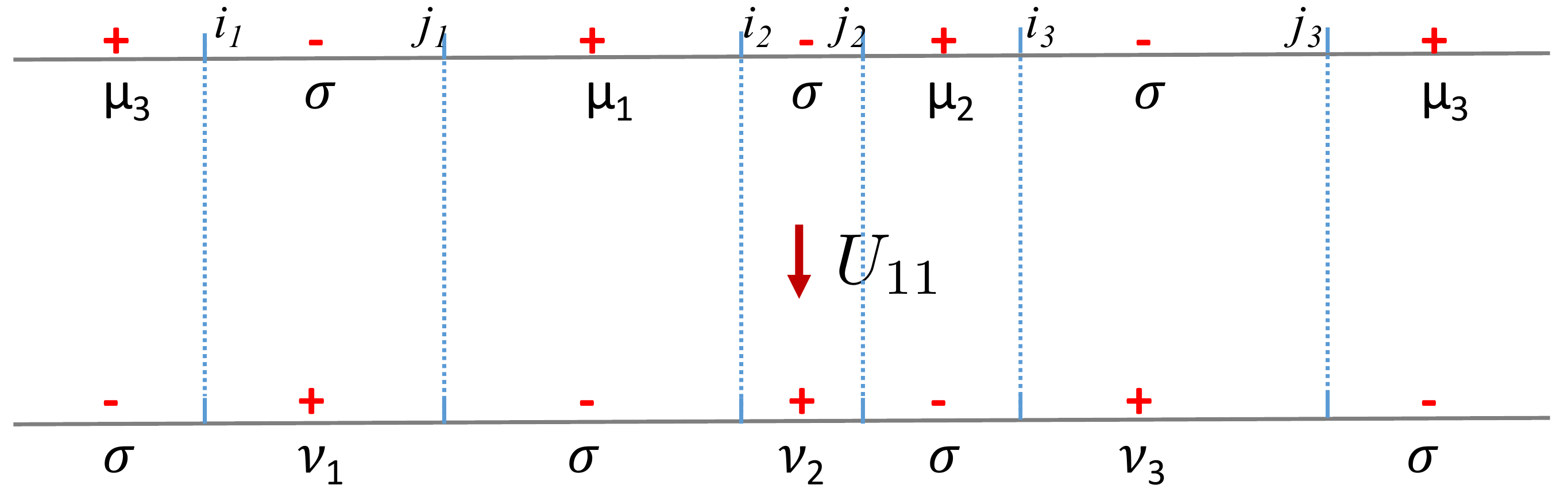}
\caption{The action of the $Z_2$ symmetry $U_{11}$ for NS spin structure and $(-1)^{\cal F} = 1$. We illustrate the case with $N_d = 3$ $+/-$ domains. The line is periodic. The consecutive $``-"$ domains in the initial state stretch from $i_k$ to $j_k$, $k = 1 \ldots N_d$.  $\mu_k/\nu_k \in \{0, 1\}$ label the $``+"$ domains in the initial/final states as either $1$ ($\mu = 0$) or $f$ ($\mu = 1$). The final state is a sum over all $\{\nu\}$ with coefficients given by Eq.~(\ref{U11}). }
\label{fig:U11}
\end{figure}

\begin{figure}[t]
\includegraphics[width = \linewidth]{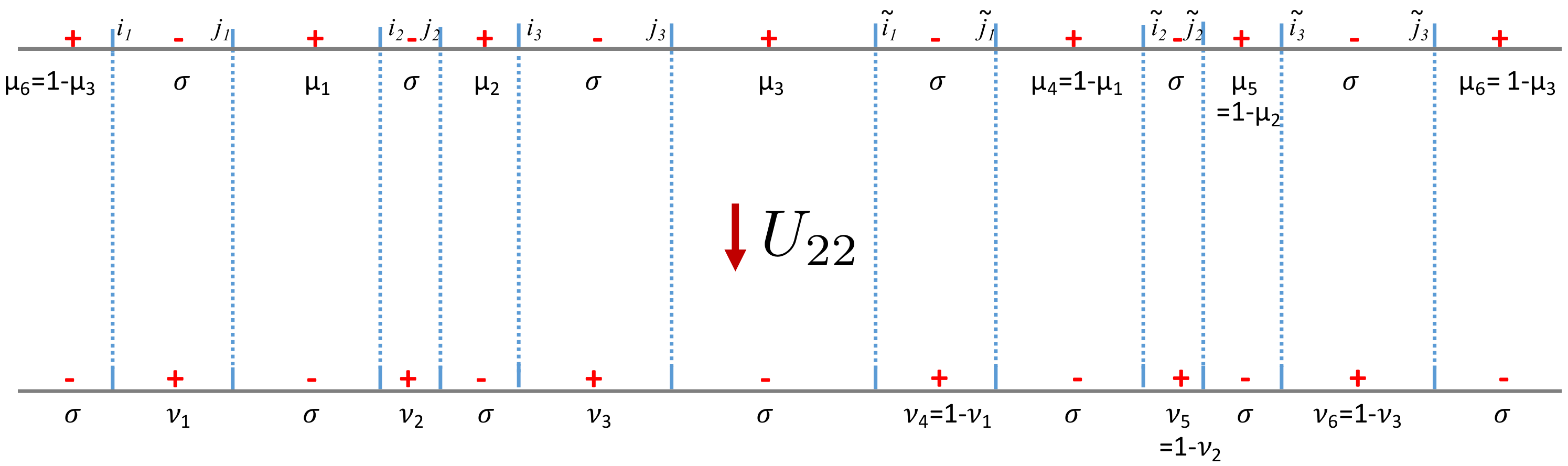}
\caption{The action of the $Z_2$ symmetry $U_{22}$ for NS spin structure and $(-1)^{\cal F} =  -1$. We illustrate the case with $N_d = 3$ $+/-$ domains. We work on the double-cover of the circle.  The consecutive $``-"$ domains in the initial state stretch from $i_k$ to $j_k$, $k = 1 \ldots N_d$. On the double-cover, we, thus, also have $``-"$ domains stretching from $\tilde{i}_k = i_k + N$ to $\tilde{j}_k = j_k + N$, $k = 1 \ldots N_d$. The consecutive $``+"$ domains in the initial/final states on the double cover are labelled by $\mu_k$/$\nu_k \in \{0,1\}$, $k  =1 \ldots 2N_d$,  with $\mu_{k+N_d} = 1-\mu_{k+N_d}$, $\nu_{k+N_d} = 1-\nu_{k+N_d}$. The final state is a sum over all $\{\nu\}$ with coefficients given by Eq.~(\ref{U22}). }
\label{fig:U22}
\end{figure}

\begin{figure}[t]
\includegraphics[width = \linewidth]{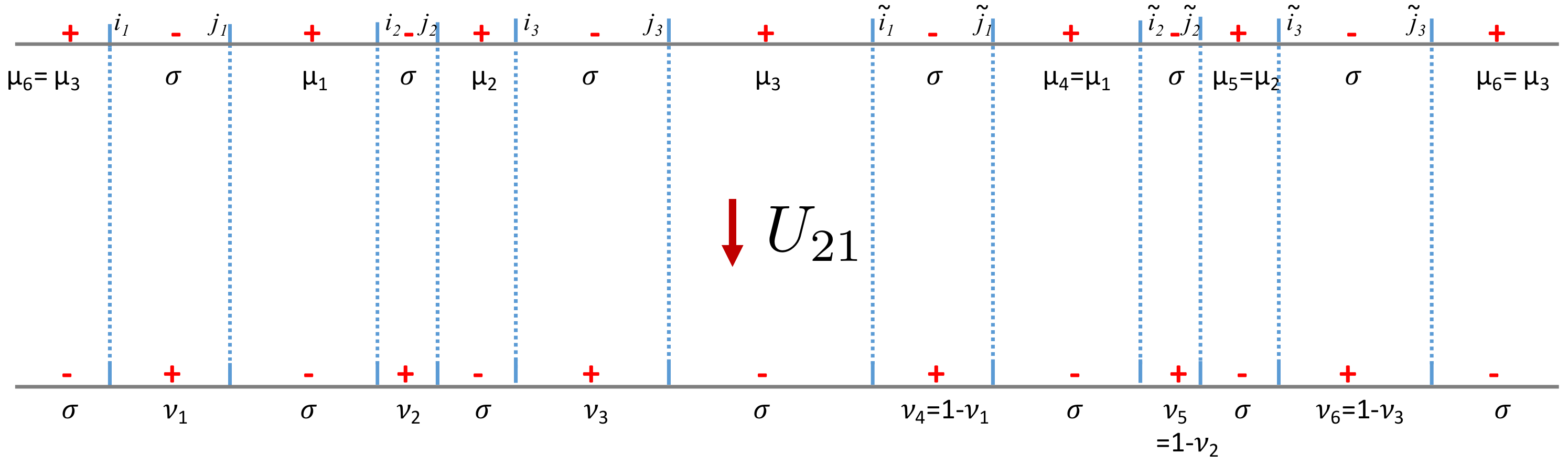}
\caption{The action of the $Z_2$ symmetry $U$ for R spin structure. In this case $U$ changes the fermion parity.  We illustrate the case where the initial state has $(-1)^{\cal F} = 1$ (no $S$-flux) and final state has $(-1)^{\cal F} = -1$ (finite $S$-flux), see Fig.~\ref{fig:NSR}. We utilize the same notation as in Fig.~\ref{fig:U22}, except now $\mu_{k + N_d} = \mu_k$ and $\nu_{k+N_d} = 1 - \nu_k$. The final state is a sum over all $\{\nu\}$ with coefficients given by Eq.~(\ref{U21}).  }
\label{fig:U21}
\end{figure}

Now, we discuss the action of the $Z_2$ ``self-duality" symmetry $U$. It has pieces that mix the $S$-flux  sectors $1$ and $2$. In block diagonal notation we write,
\beq U = \left(\begin{array}{cc} U_{11} & U_{12}\\U_{21} &U_{22} \end{array}\right) \label{Umatrix} \eeq
$U$  flips domains $``+" \leftrightarrow ``-"$. The state in the Majorana subspace, however, remains the same. We used to represent the state by grouping the Majoranas on the $``-"$ domains into complex fermions. However, the new $``-"$ domains are the old $``+"$ domains. Thus, we must perform a basis change from grouping Majoranas on the (old) $``-"$ domains to grouping them on the (old) $``+"$ domains. This results in a state which is a superposition of all possible complex-fermion occupation numbers (modulo the total fermion parity constraint) with non-trivial phase factors. Translating this to our notation for the states, we get a linear superposition of all possible $1, \sigma, f$ strings where the old $``+"$ domains turn into $\sigma$ and $``-"$ domains turn into $1$ or $f$. 
For instance, in the $1$ sector,
\beq U_{11}|\mu \rangle = \sum_{\{\nu\}} \langle \nu | U_{11}|\mu\rangle |\nu\rangle\eeq
where the matrix element
\beq \langle \nu | U_{11}|\mu\rangle  = 2^{-(N_d + 1)/2} (-1)^{\sum_{i=1}^{N_d} \mu_i (\nu_{i+1} - \nu_{i})} \label{U11}\eeq
Here, $N_d$ is the number of $``+"$ domains, which equals the number of $``-"$ domains. $\mu_i/\nu_i \in \{0, 1\}$, $i = 1 \ldots N_d$,  label the  consecutive $``+"$ domains in the initial/final state as either $1$ ($\mu = 0$) or $f$ ($\mu = 1$). See figure \ref{fig:U11} for illustration. We can also convert this to the double-cover notation, which will be necessary in other sectors
\beq \langle \nu \nu | U_{11}|\mu \mu\rangle  = 2^{-(N_d + 1)/2} i^{\sum_{i=1}^{2N_d} \mu_i (\nu_{i+1} - \nu_{i})} \label{U11double}\eeq
where $\mu_{i + N_d} = \mu_i$ and $\nu_{i + N_d} = \nu_i$. 

Likewise, in the $2$ sector,
\beq U_{22}|\mu \bar{\mu} \rangle = \sum_{\{\nu\}} \langle \nu \bar{\nu} | U_{22}|\mu \bar{\mu}\rangle |\nu \bar{\nu} \rangle\eeq
\beq \langle \nu \bar{\nu} | U_{22}|\mu \bar{\mu} \rangle  = 2^{-(N_d + 1)/2} (i)^{\sum_{i=1}^{2 N_d} \mu_i (\nu_{i+1} - \nu_{i})} \label{U22}\eeq
where $\mu_{i + N_d} = 1-\mu_i$ and $\nu_{i + N_d} = 1 - \nu_i$, see Fig.~\ref{fig:U22}.  It is not hard to see that the matrix element of $U_{22}$ is always purely imaginary. The choice of $i$ vs $-i$ in the definition (\ref{U22}) determines whether we are dealing with the phase $n = 1$ or $n = -1$ in the $Z_8$ classification (i.e. whether the right or the left mover is charged under $Z_2$ in the Ising CFT describing the edge). 

Finally, for the terms that interchange the $1$ and $2$ sectors 
\beq U_{21} |\mu \mu \rangle = \sum_{\{\nu\}} \langle \nu \bar{\nu} | U_{21}|\mu \mu\rangle |\nu \bar{\nu} \rangle \eeq
\beq \langle \nu \bar{\nu} | U_{21}|\mu \mu \rangle  = 2^{-(N_d + 1)/2} e^{-\pi i/4} (i)^{\sum_{i=1}^{2 N_d} \mu_i [\nu_{i+1} - \nu_{i}]_2} \label{U21}\eeq
where $\mu_{i + N_d} = \mu_i$ and $\nu_{i + N_d} = 1-\nu_i$, see Fig.~\ref{fig:U21}. Here $[x]_2 = 0$ if $x$ is even and $[x]_2 = 1$ if $x$ is odd. The matrix elements of $e^{\pi i/4} U_{21}$ are real. Also, $U_{21}= U^{\dagger}_{12}$. 

It is easy to check that
\bea  \langle \nu \nu | U_{11}|\mu \mu \rangle &=& \langle \nu \nu | U_{11}|\bar{\mu} \bar{\mu} \rangle = \langle \bar{\nu} \bar{\nu} | U_{11}|\mu \mu \rangle\nn\\
 \langle \nu \bar{\nu} | U_{22}|\mu \bar{\mu} \rangle &=& -\langle \nu \bar{\nu} | U_{22}|\bar{\mu} {\mu} \rangle = -\langle \bar{\nu} {\nu} | U_{22}|\mu \bar{\mu} \rangle \nn\\
  \langle \nu \bar{\nu} | U_{21}|\mu \mu \rangle &=& -\langle \nu \bar{\nu} | U_{21}|\bar{\mu} \bar{\mu} \rangle = \langle \bar{\nu} {\nu} | U_{21}|\mu \mu \rangle
\eea
This means that if we further subdivide $U$ into a block-matrix acting on sectors $1$ and $2$ and on states with $S  =\pm 1$ within each sector,
\beq U = \left(\begin{array}{cccc} U_{11} & 0 & 0 &0 \\ 0&0&U_{12} &0\\0&U_{21}&0&0\\ 0& 0&0&U_{22} \end{array}\right) \label{U44}\eeq
where the $(1,0,0,0)$  and $(0,1,0,0)$ blocks respectively denote $S  =+1$ (NS, $(-1)^{\cal F} = 1$) and $S = -1$ (R, $(-1)^{\cal F} = 1$) in the sector with no $S$-twist, and $(0,0,1,0)$ and $(0,0,0,1)$ blocks respectively denote $S = +1$ (R, $(-1)^{\cal F} = -1$) and $S = -1$  (NS, $(-1)^{\cal F} = -1$) in the sector with $S$-twist. Thus, using the correspondence in Fig.~\ref{fig:NSR},  $U: {\rm NS} \to {\rm NS}$ and $U: {\rm R} \to {\rm R}$, however, $U$ preserves fermion parity in the NS sector, but inverts fermion parity in the R sector. This is exactly what we expect. Indeed, consider the field theory (\ref{HIsing}). In the Ramond sector, we have two Majorana zero modes with momentum zero: $\chi_{R,0}$ and $\chi_{L,0}$. The fermion parity associated with these zero modes is $(-1)^{{\cal F}_0} = i \chi_{R, 0} \chi_{L,0}$, which is odd under the $Z_2$ symmetry (\ref{Z2intro}). Also, it is easy to see that the contribution of finite energy modes to fermion parity is invariant under $Z_2$. Thus, in the R sector $\{U, (-1)^{\cal F}\} = 0$, as we found.

It is easy to check that $U^2  =1$ and $U^{\dagger} = U$, as necessary.

\subsection{Hamiltonian, order parameters, fermion operators}

We now present a simple Hamiltonian obeying the $Z_2$ ``self-duality" symmetry $U$ defined in section \ref{sec:sym} (as well as the symmetry $S$ in Eq.~(\ref{S}), which must be obeyed by all local bosonic operators in the original fermion theory),
 \beq H = - \sum_i H_i \label{H3sum}\eeq
$H$ consists of local three-site terms $H_i$. $H_i$ flips the spin on site $i$ with an amplitude that depends  on the states at sites $i-1$ and $i+1$:
\bea &&H_i |\mu, \mu, \mu\rangle = c_1 |\mu, \sigma, \mu\rangle \nn\\
&&H_i |\mu, \mu, \sigma \rangle = c_2 |\mu, \sigma, \sigma\rangle \nn\\
&&H_i |\mu, \sigma, \nu\rangle = c_3 \delta_{\mu \nu} |\mu, \mu, \mu\rangle \nn\\
&&H_i |\mu, \sigma, \sigma\rangle = c_4 |\mu, \mu, \sigma\rangle \nn\\
&&H_i|\sigma, \mu, \mu\rangle = c_5 |\sigma, \sigma, \mu\rangle \nn\\
&&H_i |\sigma, \mu, \sigma\rangle  =\frac{c_6}{\sqrt{2}} |\sigma, \sigma, \sigma\rangle \nn\\
&&H_i |\sigma, \sigma, \mu\rangle = c_7 |\sigma, \mu, \mu\rangle \nn\\
&&H_i |\sigma, \sigma, \sigma\rangle =  \frac{c_8}{\sqrt{2}}(|\sigma,1, \sigma\rangle + |\sigma,f, \sigma\rangle) \nn\\
\label{H3}\eea
Here, we've indicated the states at $i-1$, $i$ and $i+1$, and $\mu$, $\nu$ stand for $1$ or $f$. The $Z_2$ symmetry and Hermiticity impose the following conditions on the eight coefficients:
\beq c_1 = c^*_3 = c^*_6 = c_8, \quad c_2 = c^*_4 = c^*_5 = c_7\eeq
Thus, there are only two independent coefficients $c_1$ and $c_2$. $c_1$ parametrizes the  amplitude for creating/annihilating a $``-"$ domain, as well as  splitting/joining two $``-"$ domains. $c_2$ parametrizes the amplitude for domain wall motion.   

We note that the Hamiltonian (\ref{H3sum}) is completely local. Moreover, if we disregard the symmetry $U$, we can impose the ``no $1$ adjacent to $f$" constraint as an energetic penalty in the Hamiltonian, rather than as a hard constraint on the Hilbert space. Then (\ref{H3sum}) becomes a regular bosonic Ising model with the Ising symmetry $S$. Let us define
\beq m_i = 1 - 2 P^{\sigma}_i \label{mi}\eeq
where $P^{a}_i$ for $a \in \{1, \sigma, f\}$ is the projector onto state $a$ on site $i$. $m_i$ is $+1$ on $``+"$ domains and $-1$ on $``-"$ domains, so it is an order parameter for the self-duality symmetry $U$. Let us break $U$ by adding 
\beq \Delta H = m \sum_i m_i\eeq 
 to the Hamiltonian. For $m \to -\infty$, in the sector with no $S$-flux, we have two degenerate ground-states $|11\ldots1\rangle$ and $|ff\ldots f\rangle$ and the Ising symmetry $S$ is spontaneously broken. For $m \to +\infty$, we have a single ground state $|\sigma \sigma \ldots \sigma\rangle$ and the Ising symmetry $S$ is restored. The point $m = 0$ is the phase-transition between these two phases. We can define an order parameter for $S$:
\beq I_i = P^{1}_i - P^{f}_i \label{Ii}\eeq 
As in the usual transverse field Ising model, this order parameter is not a local operator in the fermion theory, however, it is meaningful if we take the bosonic Ising model viewpoint.

We can also represent the fermion operators in the bosonized language. As usual, these become non-local ``string" operators:
\beq \Gamma^{\pm}_{i, i+1} = \left((-1)^{a_i} P^+_i P^-_{i+1} \pm i (-1)^{a_{i+1}} P^-_i P^+_{i+1}\right) \prod_{k = 1}^{N} S_{i+k} \label{Gammapm}\eeq
 Here, we are utilizing a double-cover notation where bonds are numbered from $1$ to $2 N$. $P^{\pm}_i$ are projectors onto $``+"$, $``-"$ Ising spin states: $P^+_i = P^{1}_i + P^{f}_i$, $P^-_i = P^{\sigma}_i$. These projectors enforce the presence of a domain wall between bonds $i$ and $i+1$, so that there is a Majorana at that location. $a_i$ here is the operator that reads off the $a$ value at position $i$; because of the projectors, $a_i \in \{0,1\}$ in the first term and $a_{i+1} \in \{0,1\}$ in the second term in parenthesis. The last term is a ``string" operator: it exchanges sectors with and without $S$-flux. $\Gamma^{\pm}_{i,i+1}$ also flips the $S$ charge. Thus, from Fig.~\ref{fig:NSR} we see that $\Gamma^{\pm}$ preserves the spin-structure (NS/R), but flips the fermion parity, as expected for a fermion operator. We note that $\Gamma^{\pm}$ has the same form (\ref{Gammapm}) in both $S$-flux sectors. Also, $\Gamma^{\pm}_{i, i+1}$ is Hermitian. One can check that under the self-duality symmetry:
\beq U\Gamma^{\pm}_{i, i+1} U^{\dagger} = \pm \Gamma^{\pm}_{i, i+1} \label{GammaU}\eeq

\subsection{Numerical Results}
\label{sec:numerics}
\begin{figure}
\includegraphics[width = \linewidth]{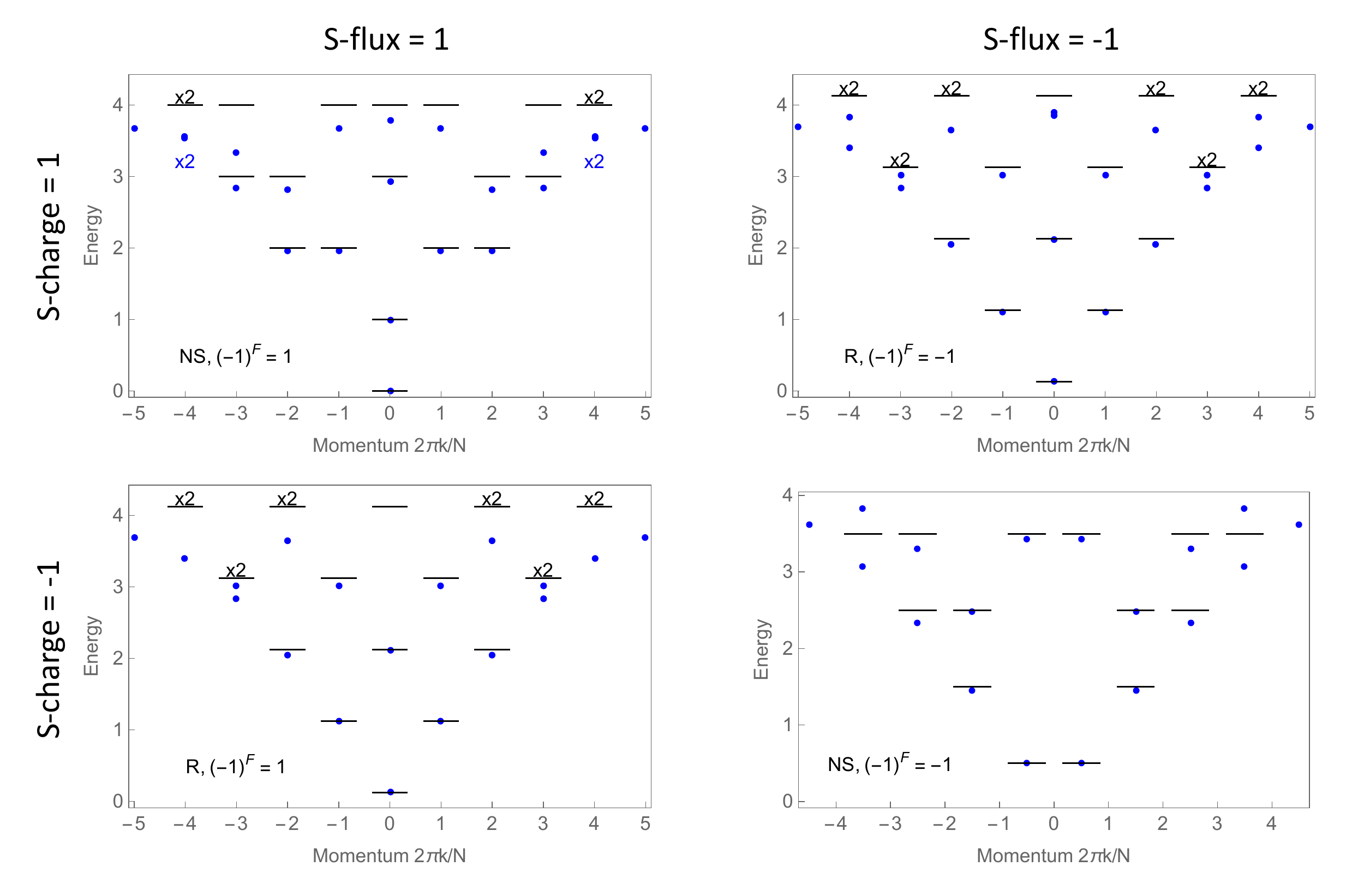}
\caption{ Exact diagonalization of the 1d Hamiltonian (\ref{H3}) for a chain of length $N  =16$ with periodic boundary conditions.
Top left and bottom right plots correspond to NS sector with even and odd fermion parity respectively. Top right and bottom left plots correspond to R sectors with opposite fermion parity. The energy difference between the ground state and the first excited state, $E_1 - E_0$, in the NS sector with $(-1)^{\cal F} = 1$ is normalized to $1$. Momentum is measured in units of $2\pi/N$. Horizontal dashes indicate the predictions of the Ising CFT.}
\label{fig:ED16}
\end{figure}

\begin{figure}
\includegraphics[width = \linewidth]{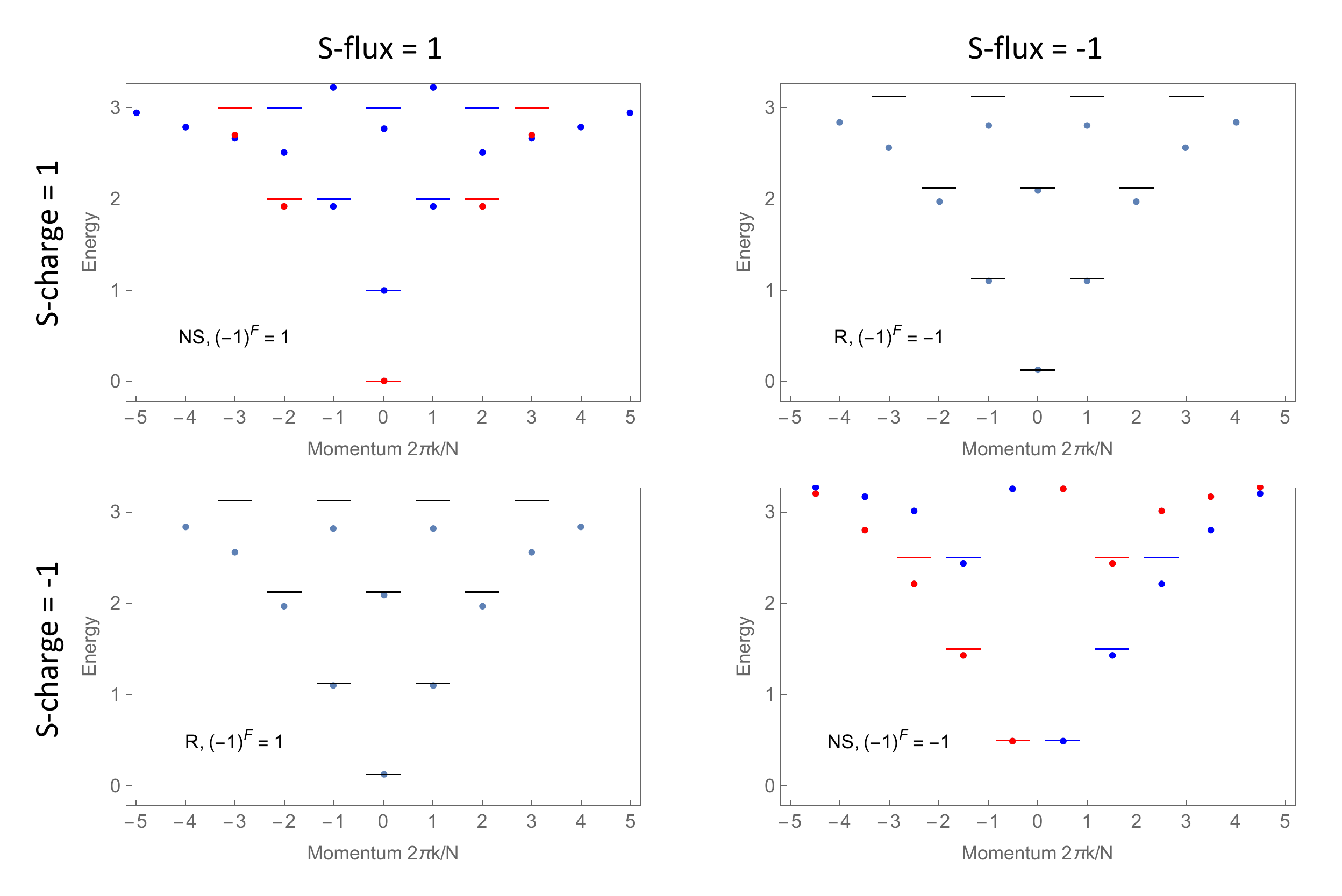}
\caption{Same as Fig.~\ref{fig:ED16}, except for $N = 12$. In the NS sectors, the quantum number under the $Z_2$ self-duality symmetry $U$ (\ref{Umatrix}) is marked as red ($U = +1$) or blue ($U = -1$). The R sectors with opposite fermion parity are interchanged by $U$ and have the same spectrum.}
\label{fig:ED12}
\end{figure}

{\it Exact diagonalization.} We have performed exact diagonalization on the Hamiltonian (\ref{H3}) with $c_1  = c_2 >0$ for  up to $N = 16$ sites arranged on a circle. The low-lying spectrum for $N = 16$ together with the predictions from the Ising CFT is displayed in Fig.~\ref{fig:ED16}. The four plots correspond to NS and R sectors with $(-1)^{\cal F}  =\pm 1$, which are studied via the prescription in Fig.~\ref{fig:NSR}. 40 top eigenvalues are kept in each $S$-flux sector. We find good agreement with the CFT predictions. In particular, the spectra in the $R$ sector with $(-1)^{\cal F} =  1$  and $(-1)^{\cal F} =  -1$ are identical, as these sectors are interchanged by the $Z_2$ self-duality symmetry $U$. We also study the quantum numbers under $U$ in the NS sector; for numerical reasons we were only able to do this for slightly smaller systems up to $N = 12$. Our findings are shown in Fig.~\ref{fig:ED12} - again, we find good agreement with CFT predictions.

\vspace{0.5cm}

{\it DMRG.} We have also performed DMRG studies of the Hamiltonian (\ref{H3}) with parameters $c_1 = c_2 > 0$ using the iTensor library.\cite{ITensor} We study chains of length up to $N = 500$ with open boundary conditions. More precisely, the sites $0$ and $N+1$ are taken to be in state $1$; as we have discussed in section \ref{sec:Hilbert} this corresponds to boundary conditions which break the $Z_2$ self-duality symmetry - both boundary Ising spins are ``+". Furthermore, this is the sector where the fermion parity of the segment is even. Also, if we take the viewpoint of the bosonic Ising model, choosing the boundary spin to be $1$ rather than $f$ breaks the ``Ising" symmetry $S$.

\begin{figure}[t]
\includegraphics[width = 0.45\linewidth]{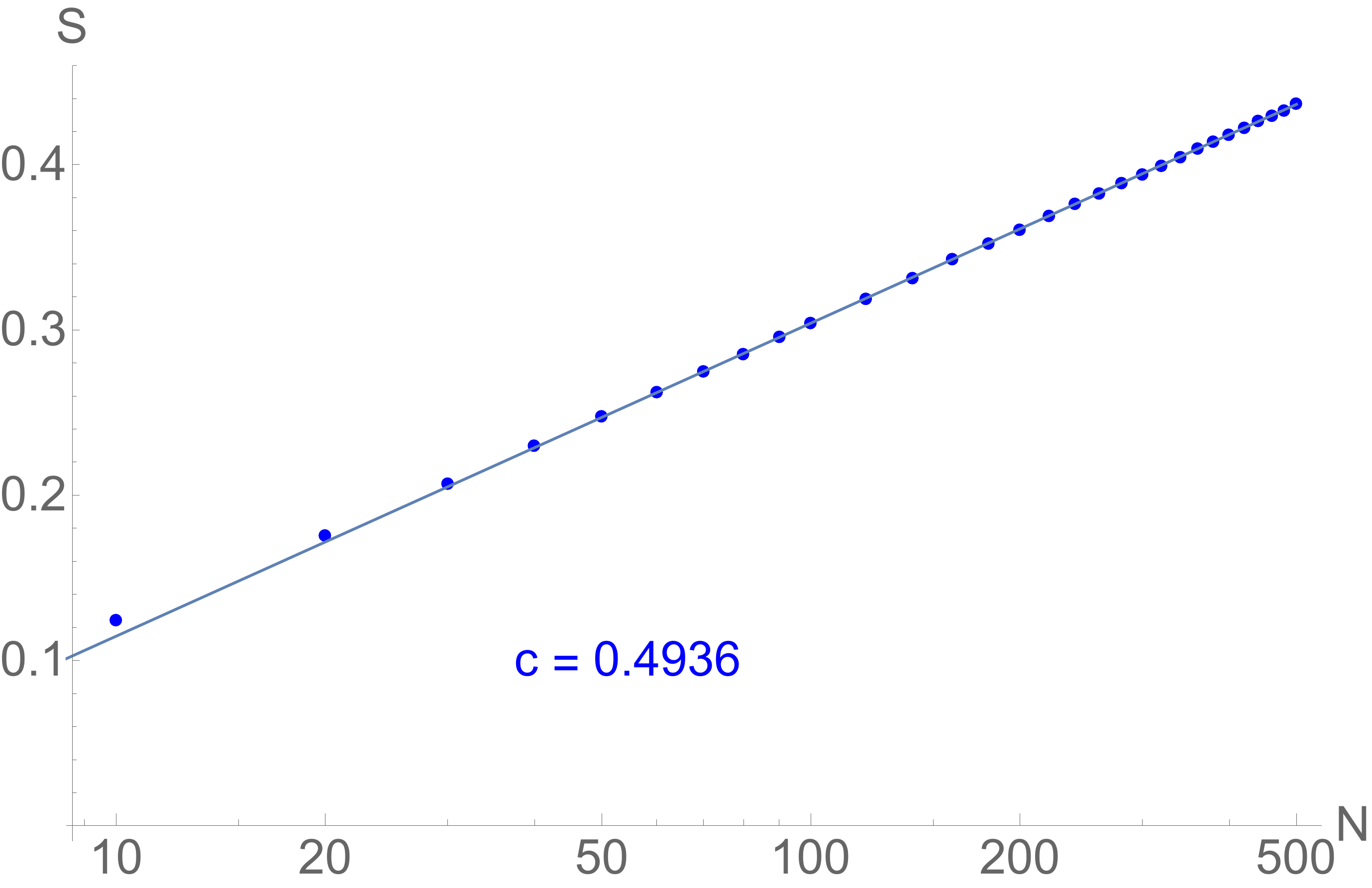} \quad \includegraphics[width = 0.45\linewidth]{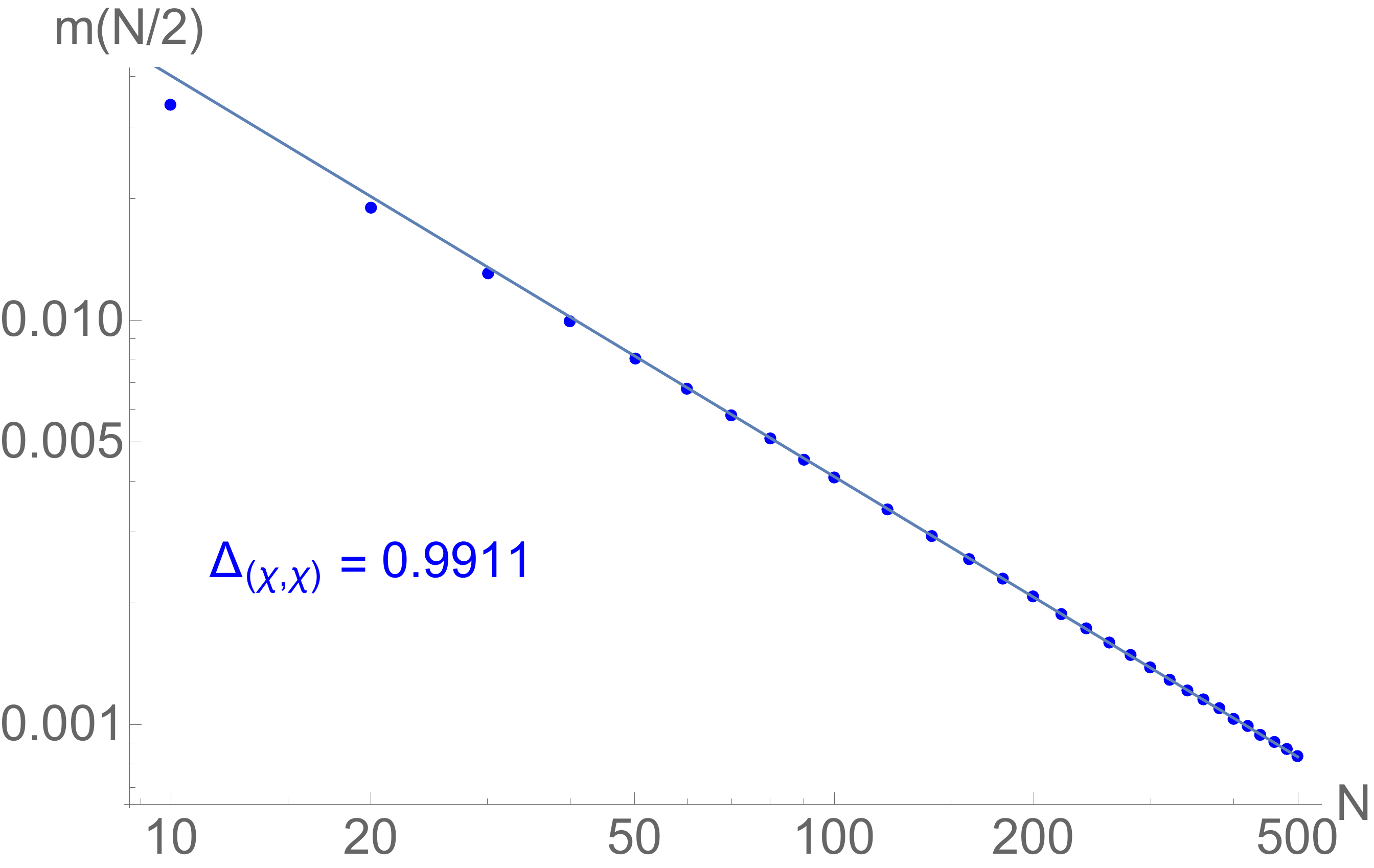}

\vspace{1cm}

\includegraphics[width = 0.45\linewidth]{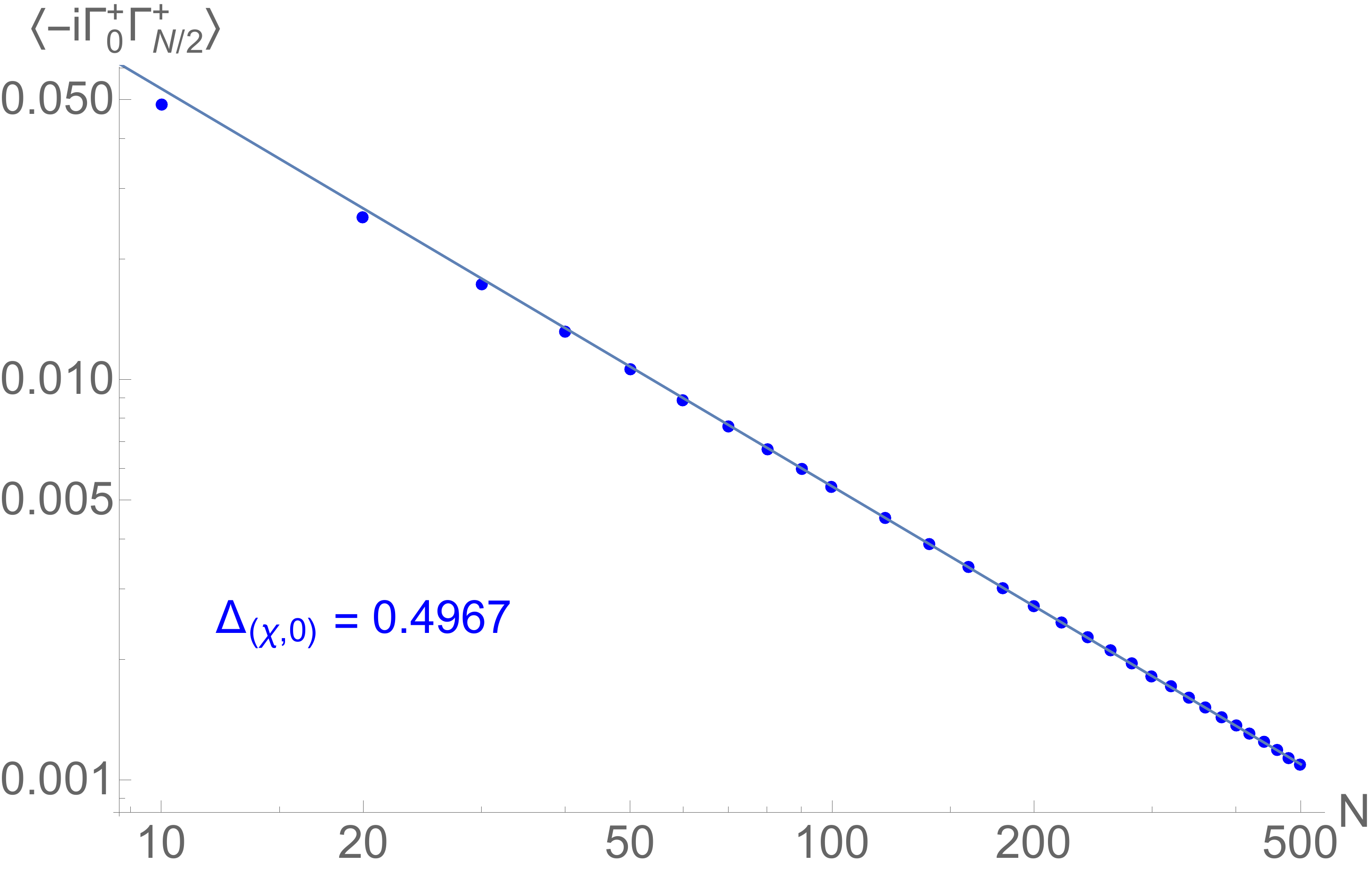} \quad \includegraphics[width = 0.45\linewidth]{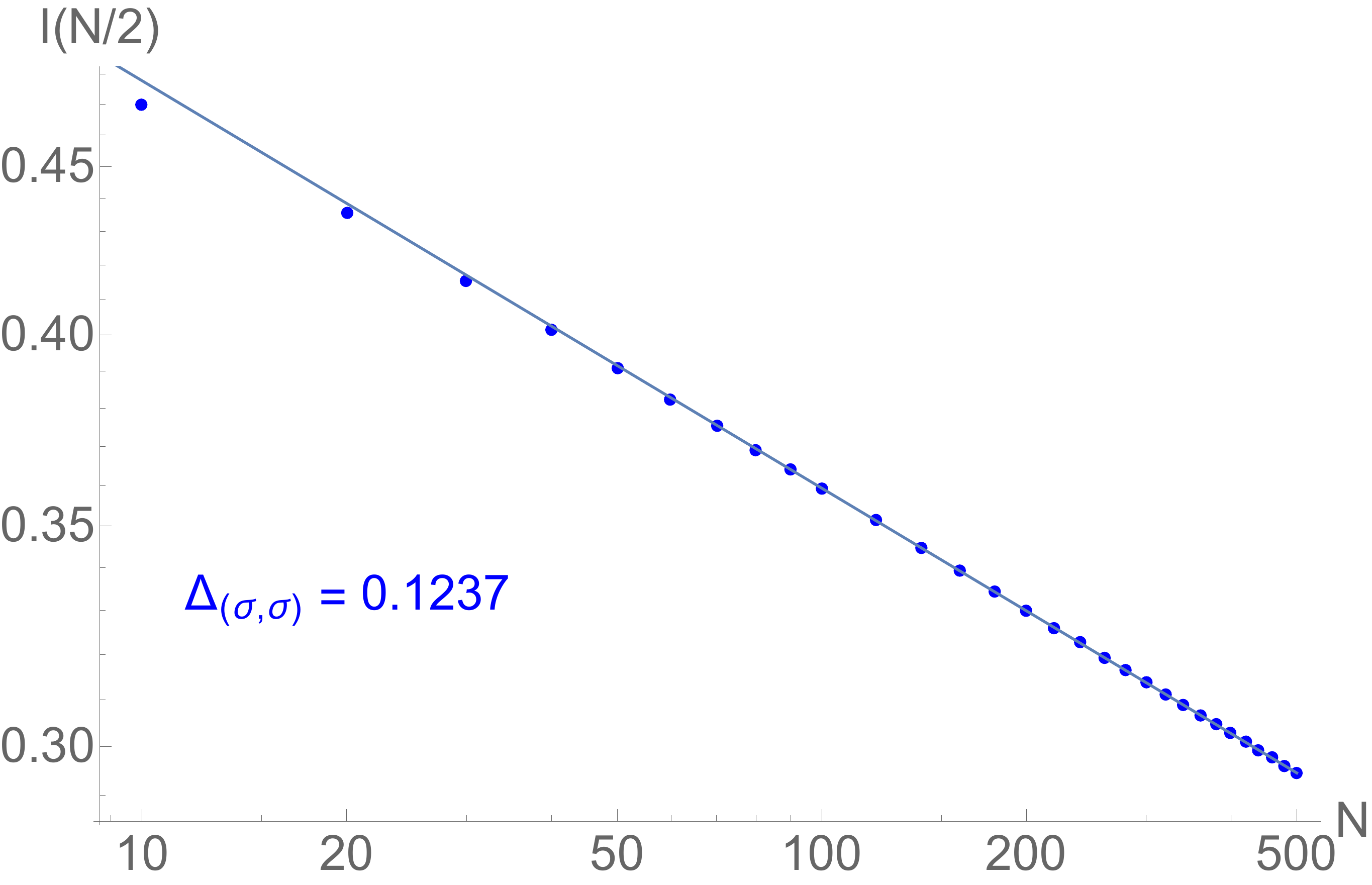}
\caption{DMRG simulations of the 1d Hamiltonian (\ref{H3}) for a chain of length $N$ with open boundary conditions. All fits are performed for the range $N = 100\ldots 500$. Top left: the entanglement entropy $S$ for the chain cut at the center. Top, right: the order parameter for the $Z_2$ self-duality symmetry, Eq.~(\ref{mi}), $\langle m(x)\rangle  \sim \langle i \chi_R \chi_L(x) \rangle$,  at $x = N/2$.  Bottom left: two point function $\langle \Gamma^+_{0,1} \Gamma^+_{N/2,N/2+1}\rangle$ of  fermion operator $\Gamma^+$ in Eq. (\ref{Gammapm}); $\Gamma^+(x) \sim \chi_R(x)$. Bottom, right: the order paramer for the $Z_2$ ``Ising" symmetry $S$, Eq.~(\ref{Ii}), $\langle I(x) \rangle \sim \langle \sigma_R \sigma_L(x) \rangle$, at $x = N/2$.  }
\label{fig:DMRG}
\end{figure}

Our DMRG findings are shown in Fig.~\ref{fig:DMRG}. The top, left of Fig.~\ref{fig:DMRG} displays the entanglement entropy $S$ for the open chain cut at the center. In a CFT $S$ should obey, $S = \frac{c}{6} \log N + const$, with $c$ - the central charge. We extract $c \approx 0.4936$, consistent with the Ising CFT value $c = 1/2$. 

The top, right of Fig.~\ref{fig:DMRG} displays the expectation value of the order parameter for the $Z_2$ self-duality symmetry (\ref{mi}), $\langle m_i \rangle$, at the center of the chain $i  =N/2$. 
In the continuum theory (\ref{HIsing}), $m(x) \sim i \chi_R \chi_L(x)$. Further, since our boundary conditions break the $Z_2$ self-duality symmetry, we expect $\langle m(x) \rangle$ to be non-zero and to obey the scaling form $\langle m(x) \rangle = \frac{1}{N^{\Delta(\chi,\chi)}} f(x/N)$, where $\Delta(\chi, \chi) = 1$ is the scaling dimension of  $i \chi_R \chi_L$. We extract $\Delta(\chi, \chi) \approx 0.9911$. 

The bottom, left of Fig.~\ref{fig:DMRG}  displays the two-point function $\langle \Gamma^+_{0,1} \Gamma^+_{N/2,N/2+1}\rangle$ of fermion operators $\Gamma^+$ (\ref{Gammapm}). Given the symmetry properties (\ref{GammaU}) we expect $\Gamma^+(x) \sim \chi_L(x)$ and $\Gamma^-(x) \sim \chi_R(x)$. An explicit calculation shows that with this boundary conditions $\langle i \chi_L(x) \chi_L(y)) \rangle \sim \frac{1}{N \sin(\pi(x-y)/2N)}$, i.e. the boundary and bulk scaling dimension of $\chi_L$ is $\Delta(\chi,0) = 1/2$. We extract $\Delta(\chi,0) \approx 0.4967$.

The bottom, right of Fig.~\ref{fig:DMRG} displays the expectation value of the order parameter for the $Z_2$ Ising symmetry $S$, $\langle I_i \rangle$, Eq.~(\ref{Ii}), at $i =N/2$. 
In the continuum theory, $I(x) \sim \sigma_L \sigma_R(x)$, where $\sigma_{L/R}$ are operators, which twist the phase of $\chi_{L/R}$ by $\pi$. Since our boundary conditions break the Ising symmetry $S$, we expect the scaling form $\langle I(x) \rangle = \frac{1}{N^{\Delta(\sigma,\sigma)}} g(x/N)$, where $\Delta(\sigma, \sigma) = 1/8$ is the scaling dimension of the $\sigma_L \sigma_R$ operator. We extract $\Delta(\sigma, \sigma) \approx 0.1237$.

\subsection{Odds and ends}
\label{sec:odds}

{\it Other odd $\nu$.} Recall that 2d $Z_2 \times Z^f_2$ SPTs have a $Z_8$ classification. Above we have focused on the case with $n = 1$. We can easily adapt our 1d boundary model to the case of other odd $\nu$. For $n = -1 \sim 7$, we simply replace the symmetry action $U$ in Eq.~(\ref{Umatrix}) by $U^*$. What about $n = 3$ and $n = 5$? We recall that the boundary of the effectively bosonic $n = 4$ phase can be mimicked with a simple tensor product Hilbert space consisting of spin $1/2$ per site and a symmetry action
\beq U_{n =4} = (-1)^{N_{dw}/2} \prod_i \sigma^x_i \eeq
where $N_{dw}$ is the number of domain walls.\cite{Levin_Gu} Thus, returning to our odd $\nu$ phases, we take
\beq U_{n + 4} = (-1)^{N_{dw}/2} U_n\eeq
where $N_{dw}$ is the number of domain walls between our ``Ising" spins.

\vspace{0.3cm}

{\it General groups $G \times Z^f_2$.} We recall that for a general symmetry group $G \times Z^f_2$ one can generate all 2d SPTs in the following way.\cite{FidkowskiSpin} Pick a homomorphism $\mu: G \to Z_2$. Consider a system where symmetry $G$ acts on states in a $Z_2$ fashion via $\mu$, i.e. all fermions transform in representation $\mu$ of $G$ or in a trivial representation. Now viewing $G$ as a $Z_2$ symmetry via $\mu$, build a $Z_2 \times Z^f_2$ SPT out of these fermions. All other SPTs can be obtained by stacking a supercohomology $G \times Z^f_2$ SPT on top. Thus, all beyond supercohomology SPTs with symmetry group $G \times Z^f_2$ effectively reduce to a $Z_2 \times Z^f_2$ SPT, and we can use our 1d model for the boundary.

\vspace{0.3cm}

{\it Time-reversal ${\cal T}$ with ${\cal T}^2 = (-1)^{\cal F}$.}  We now discuss the symmetry group $Z^T_4$  generated by the anti-unitary time-reversal symmetry ${\cal T}$, which satisfies ${\cal T}^2 = (-1)^{\cal F}$. This is the symmetry of superconductors with spin-orbit coupling where time-reversal acts on spinfull electrons $c_{\sigma}$ via,  ${\cal T}: c_{\uparrow} \to c_{\downarrow}, \, c_\downarrow \to - c_\uparrow$. For non-interacting electrons, this symmetry class is known as DIII. In 2d, non-interacting phases in this class have a $Z_2$ classification.\cite{KitaevNI, LudwigNI, LudwigNI2} Interactions don't alter this $Z_2$ classification.\cite{KapustinFerm,FreedHopkins} The non-trivial phase can be obtained by putting the spin-up electrons into a $p+ip$ superconductor and spin-down electrons into a $p-ip$ superconductor. In this construction, the edge is again described by the Majorana CFT (\ref{HIsing}), where ${\cal T}$ acts via
\beq {\cal T}: \chi_R \to \chi_L, \quad \chi_L \to - \chi_R \label{chiT} \eeq
We note that this phase can also be realized with a  commuting projector bulk Hamiltonian, see Ref.~\onlinecite{ChenT2d}. While we have not explicitly derived an effective 1d edge model starting from the bulk Hamiltonian in Ref.~\onlinecite{ChenT2d}, it is easy to guess how to adapt our 1d model above for this purpose. In fact, one can use exactly the same effective 1d boundary Hilbert space, and implement ${\cal T}$ as
\beq {\cal T}  = U K \eeq
where $U$ is still given by Eqs.~(\ref{Umatrix}), (\ref{U11}), (\ref{U22}), (\ref{U21}) and $K$ is the complex conjugation operator. Then ${\cal T}$ acts on the fermion operators (\ref{Gammapm}) as
\beq {\cal T} \Gamma^+ {\cal T}^{\dagger} =  - \Gamma^-, \quad {\cal T} \Gamma^- {\cal T}^{\dagger} = \Gamma^+ \label{GammaT}\eeq
Now the Hamiltonian (\ref{H3}) with $c_1 = c_2 > 0$ that we studied numerically in section \ref{sec:numerics} is invariant under both $U$ and $K$.  Further, we saw that our numerical results were consistent with a Majorana CFT where  $\Gamma^+ \sim \chi_L$ and $\Gamma^- \sim \chi_R$, implying precisely the transformation properties (\ref{chiT}).

It is also interesting to explicitly compute ${\cal T}^2 = (UK)^2$. Starting from (\ref{U44})
\beq {\cal T}^2 = (UK)^2 = \left(\begin{array}{cccc} 1 & 0 & 0 &0 \\ 0&i&0 &0\\0&0&-i&0\\ 0& 0&0&-1 \end{array}\right) \eeq
Thus, in the NS sector ${\cal T}^2  = (-1)^{\cal F}$, while in the R sector ${\cal T}^2 = i (-1)^{\cal F}$. Thus, in the NS sector ${\cal T}$ satisfies the expected group law. On the other hand, in the R sector the  ${\cal T}$ action fractionalizes in the way precisely expected from the Majorana CFT (\ref{HIsing}) and action (\ref{chiT}): namely ${\cal T}$ anticommutes with fermion parity and ${\cal T}^2 = \pm i$, see e.g. Ref.~\onlinecite{MCFV2014}, section 5.

\section{Derivation}
\label{sec:der}
In this section we derive the effective 1d boundary model of section \ref{sec:model} starting from the TF model for the $\nu =1$ $Z_2 \times Z^f_2$ fSPT.\cite{FidkowskiSpin}

\subsection{Bulk Tarantino-Fidkowski model}

\begin{figure}[h!]
\centering
\includegraphics[width=0.8\textwidth]{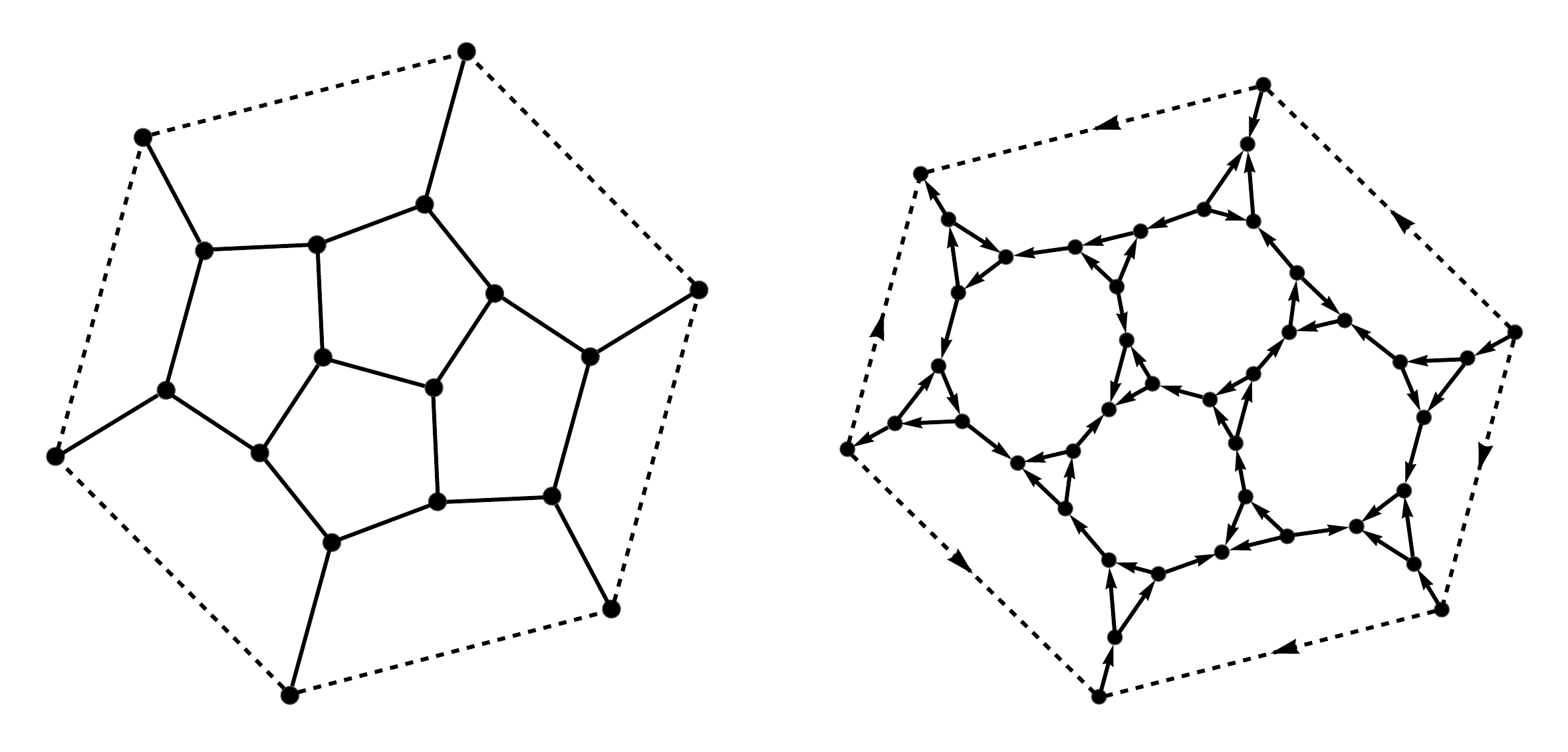}
\caption{Introducing an edge to the Tarantino-Fidkowski model. Left: graph \({\cal G}\) which illustrates ``physical'' plaquettes; Ising spins live on the faces of this graph. Boundary edges are dashed. All edges, except the boundary, carry a complex fermion. Right: graph \({\cal G}'\) where the vertices are Majorana fermions. ${\cal G}'$ is given a Kasteleyn orientation: all edges are oriented so that there is an odd number of clockwise edges around any face.
}
\label{fig:Gopen}
\end{figure}

TF begin with a trivalent graph $\cG$ embedded into a closed genus \(g\) surface (Fig.~\ref{fig:Gopen}, left, bulk). An Ising spin $\tau^a_p$ lives on each face $p$ of $\cG$. There is also a complex fermion $c_l$ living on each edge $l$ of \(\cG\). The Ising symmetry operator simply flips all the Ising spins:
\beq U = \prod_{p \in \cG} \tau^x_p \label{UTF} \eeq
Note that $U$ acts trivially on the fermion degrees of freedom.

As a next step, each vertex of \(\cG\) is blown up into a triangle to make a new graph \(\cG'\) (Fig.~\ref{fig:Gopen}, right, bulk). Edges  connecting vertices belonging to different triangles of $\cG'$ are referred to as type I. Edges connecting vertices within the same triangle are referred to as type II. The graph \(\cG'\) is also given a Kasteleyn orientation; i.e. edges are oriented in such a way that the number of clockwise-oriented edges around each face is odd (this applies to both faces derived from faces of $\cG$ and to the triangular faces). For an edge between
vertices $i$ and $j$ we let $s_{ij} = 1$ if the edge is oriented from $i$ to $j$, and $s_{ij} = -1$ otherwise. The complex fermion degrees of freedom $c_l$ that originally lived on edges $l$ of $\cG$ (i.e. on type I edges of $\cG'$) are now split into  Majorana fermions $\gamma_i$ living on vertices of $\cG'$ such that
\beq \gamma_i = (c_l + c^{\dagger}_l), \quad \gamma_j  = i (c^{\dagger}_l - c_l) \label{clsplit} \eeq
for an edge $l$ oriented from $i$ to $j$. 

Note that only faces of $\cG'$ derived from the original faces of $\cG$ carry dynamical Ising spins - we refer to these faces as plaquettes. However, we can extend the Ising spin assignments to the triangular faces $T$ of $\cG'$ using the majority rule: if the majority of the three plaquettes bordering $T$ has spin $\tau^z = \tau$  we assign $T$ to have spin $\tau$. Note that spins on triangular faces are thus completely slaved to the spins on the plaquettes.

\begin{figure}[h!]
\centering
\includegraphics[width=\textwidth]{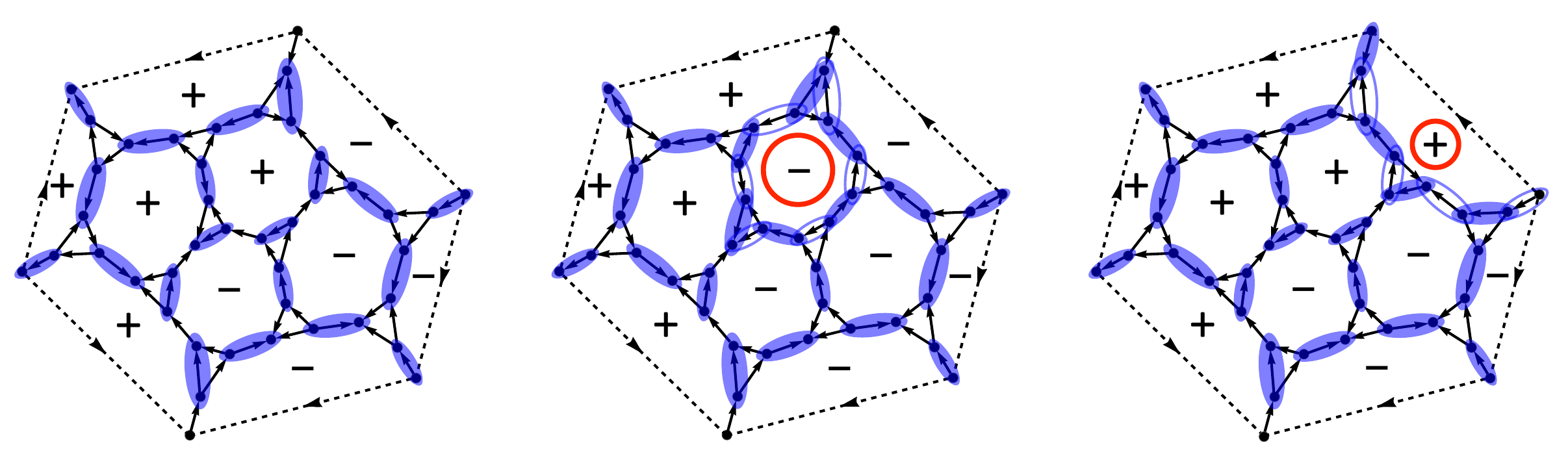}
\caption{Effect of the plaquette flip operators $F_p$. Left: Spin configuration and associated dimer cover ${\cal D}$ (filled ovals). Middle: left configuration with the bulk plaquette in red flipped  - the new dimer cover ${\cal D'}$ is shown with filled ovals.  Part of the old dimer cover $({\cal D} + {\cal D'})\cap{\cal D}$ is shown  with empty ovals. Filled and empty ovals form a closed loop $({\cal D} + {\cal D'})$ around the flipped plaquette. Right: Left configuration with boundary plaquette in red flipped. Same notation as in the middle figure. Now filled and empty ovals $({\cal D} + {\cal D'})$ form an open segment around the flipped plaquette.
} 
\label{fig:Flip}
\end{figure}

Each Ising spin configuration induces a dimer cover of \(\cG'\) as follows. A type I edge  is covered if the two faces bordering it have the same spins, while a type II edge is covered if the two faces bordering it have opposite spins (Fig. \ref{fig:Flip}, left). From now on, we work in a subspace ${\cal V}^c$ of the full Hilbert space where the Majoranas are slaved to the Ising spins according to the dimer covering: 
if an edge $ij$ is covered by a dimer then $i s_{ij} \gamma_i \gamma_j = 1$. We  enforce this constraint with a local Hamiltonian $H_{fermion}$,
\beq H_{fermion} = - \sum_{ \langle ij \rangle \in {\rm Type\, I} } (1-D_{ij}) i s_{ij} \gamma_i \gamma_j - \sum_{ \langle i j\rangle \in {\rm Type\, II}} D_{ij} i s_{ij} \gamma_i \gamma_j \label{Hferm}\eeq
Here, $D_{ij} = 0$ if the two faces bordering the edge $ij$ have the same spin, and $D_{ij} = 1$ if the two faces bordering $ij$ have opposite spins. 

For a given plaquette $p$, we define the neighbourhood that includes $p$ and all the triangles bordering it as $\d'p$. If we flip the spin on $p$ then the spins on triangles in $\d' p$ can also change. Furthermore, the dimer cover changes. If we call the old dimer cover ${\cal D}$ and the new dimer cover ${\cal D}'$ then ${\cal D}+{\cal D}'$ forms a closed loop composed of some edges of $\d' p$ (here ${\cal D}+{\cal D}'$ consists of all the edges in ${\cal D}$ or ${\cal D}'$ but not in both).  The consecutive edges in this loop alternate between dimers in ${\cal D}$ and dimers in ${\cal D}'$. We now define the plaquette flip operator $F_p$ as

\beq F_p=\sum_c X_{p,c}\otimes(\tau_p^x P_{p,c}), \label{Fpdef} \eeq
Here $c$ runs over all Ising spin configurations of $p$ and the plaquettes bordering it. The right tensor factor $(\tau_p^x P_{p,c})$ operates on the spins and first projects onto a given spin configuration $c$ and then flips the spin at \(p\).  The left tensor factor \(X_{p,c}\) operates on the Majoranas in $\d' p$ and ensures that the final state is consistent with the  dimer cover.  Let $c_p$ be the  Ising spin configuration obtained by flipping the spin \(p\) in configuration \(c\); let the dimer coverings corresponding to $c$ and $c_p$ be ${\cal D}(c)$ and ${\cal D}(c_p)$.  Further, label the consecutive sites in the loop ${\cal D}(c) + {\cal D}(c_p)$ as $1 \ldots 2 n$, with edges $(1,2), (3,4), \ldots  (2i - 1, 2i), \ldots (2n-1, 2n)$ in ${\cal D}(c)$, and edges $(2,3), (4,5), \ldots (2i, 2i+1), \ldots (2n, 1)$ in ${\cal D}(c_p)$. (The direction of the loop and the basepoint are irrelevant.) We then have:
\beq X_{p,c}=N_{p,c} P_{23} P_{45} \ldots P_{2i, 2i+1} \ldots P_{2n,1} 
 \label{Xbulk} \eeq
where $N_{p,c} = 2^{(n-1)/2}$ is a normalization factor and
 $P_{ij} = \frac12(1 + i s_{ij} \gamma_i \gamma_j)$ is a projector onto the fusion channel $i s_{ij} \gamma_i \gamma_j = 1$ of Majoranas $i$ and $j$. Thus, (\ref{Xbulk}) is up to normalization a  projector onto the new dimer cover. As TF showed,  $F_p |\psi\rangle$ has the same norm as $|\psi\rangle$, provided that the Majoranas in $|\psi\rangle$ are consistent with the dimer covering. 

As we will review shortly, $[F_p, F_q] = 0$ when acting in ${\cal V}^c$ for arbitrary plaquettes $p$, $q$. The TF Hamiltonian is simply a sum,
\beq H_{TF} = - \sum_{p} F_p + H_{fermion} \eeq

\subsection{Introducing the edge}
We now introduce an edge into the TF model, see Fig.~\ref{fig:Gopen}. We begin with an open surface  covered by a collection of faces ${\cal G}$, see Fig.~\ref{fig:Gopen}, left. We require every vertex in ${\cal G}$ (including the boundary vertices) to be trivalent. Further, no two vertices on the boundary connect to the same bulk vertex. As before, every face of ${\cal G}$ (including boundary faces) carries an Ising spin variable. Also, every edge $l$ of ${\cal G}$, {\it except the boundary edges}, carries a complex fermion $c_l$. We draw the boundary edges as dashed in Fig.~\ref{fig:Gopen}. We blow up each vertex  that is {\it not} on the boundary of ${\cal G}$ into a triangle obtaining a graph ${\cal G}'$, see Fig.~\ref{fig:Gopen}, right. 
 We again choose a Kasteleyn orientation on ${\cal G}'$ and split the complex fermions $c_l$ on (non-boundary) edges inherited from ${\cal G}$ into Majorana fermions according to (\ref{clsplit}). There is again one Majorana at every vertex of ${\cal G}'$. We again call the faces of ${\cal G}'$ inherited from the faces of ${\cal G}$ - plaquettes, and these carry dynamical Ising spins. The plaquettes on the boundary of ${\cal G}'$ are referred to as boundary plaquettes/spins, all the other plaquettes are referred to as bulk plaquettes. The Ising spin configuration can be extended to the new triangular faces of ${\cal G}'$ by the majority rule as before. We can also extend the dimer  rules to all the edges not on the boundary of ${\cal G}'$. The boundary edges of ${\cal G}'$ 
 {\it never} carry a dimer (we can label them as type III). Note that edges connecting boundary Majoranas to the bulk are treated as type I, so the dimer  rules produce an unpaired Majorana on the boundary of ${\cal G}'$ whenever there is a domain wall between boundary spins. Here and below, we use ``unpaired'' to describe Majoranas unconstrained by the dimer configuration, and ``fused'' to describe the state of the unpaired Majoranas (if it is known). On the other hand, if the Majorana is covered by a dimer, we say that it is ``paired." Again, below we only work in the Hilbert space ${\cal V}^c$ where the Majoranas conform to the dimer configuration.

We now let the bulk Hamiltonian be
\beq H_{bulk} = -\sum_{p \in bulk} F_p + H_{fermion} \label{Hbulk2} \eeq
where the first sum is over only bulk plaquettes $p$ and $H_{fermion}$, Eq.~(\ref{Hferm}), constrains the Majoranas (including boundary Majoranas) to the dimer configuration. Clearly, $H_{bulk}$ leaves the boundary spins unconstrained, so its ground state manifold consists of all boundary spin configurations, where for each fixed boundary spin configuration with $N_{dw}$ domain walls there is an additional degeneracy of $2^{N_{dw}/2}$ coming from the unpaired Majoranas. 





\begin{figure}[h!]
\centering
\includegraphics[width=0.9\textwidth]{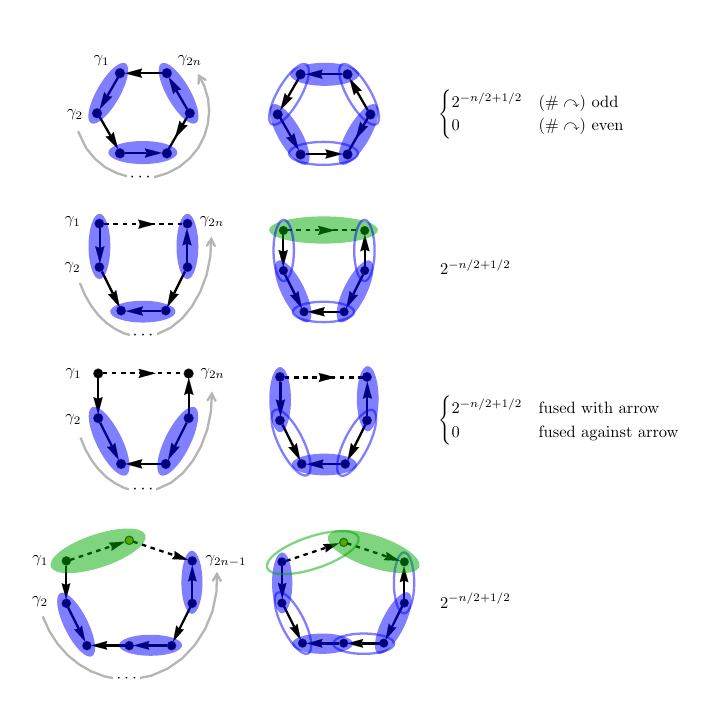}
\caption{Plaquette flip operator $F_p$: the fermion factor $X_{p,c}$. The left and right column show the initial and final configurations. Filled and empty blue ovals indicate dimers as in Fig.~\ref{fig:Flip}. The top row corresponds to bulk plaquette flips, and the rest to boundary plaquette flips.  For each row, we suppose the system is initially in the state on the left, we then apply the projectors corresponding to the blue filled ovals on the right. \newline
First row:   rotating dimers around a closed loop annihilates the state if the loop is not Kasteleyn oriented, and otherwise shrinks its norm to \(2^{-n/2+1/2}\).\newline
Second row:  A solid green dimer indicates the fusion channel of unpaired Majoranas: $i s_{1, 2n} \gamma_1 \gamma_{2n} = 1$.  This follows immediately from the first row: we can multiply the projector on the right by \(\boldsymbol{1}=\frac{1 + i s_{1,2n}\gamma_1 \gamma_{2n}}{2} + \frac{1 - i s_{1,2n} \gamma_1 \gamma_{2n}}{2}\), and after expanding, the second term is zero. Thus, the two unpaired Majoranas are fused to respect the Kasteleyn orientation and the norm is  \(2^{-n/2+1/2}\). \newline
Third row: If the fusion state of the unpaired Majoranas on the left is $i s_{1, 2n} \gamma_1 \gamma_{2n}  = -1$, the state is annihilated, otherwise, the norm is \(2^{-n/2+1/2}\).
\newline
Fourth row: There are $2n-1$ Majoranas in ${\cal D}(c)+{\cal D}(c_p)$. The gray  Majorana on top is auxilliary (e.g. another unpaired boundary Majorana from a different plaquette) and is assumed to be initially fused with $\gamma_1$ along the green oval. In the final state, it becomes fused with $\gamma_{2n-1}$.
}
\label{fig:create}
\end{figure}

To further analyze the boundary, we extend the definition of the \(F_p\) operators to the case where \(p\) is a boundary spin. Instead of closed loops of Majoranas having their dimer covering shifted by one, when a boundary spin is flipped the dimer configuration shifts along an {\it open string}. Unlike the closed loop case, shifting a dimer configuration along an open string can cause unpaired Majoranas to appear, disappear, or shift around.  We take $F_p$ to again have the form (\ref{Fpdef}), with the bosonic factor $\tau^x_p P_{p,c}$ the same as in the bulk. We modify the fermionic factor $X_{p,c}$ as follows. We have three cases to consider:

\begin{itemize}
\item Both boundary Majoranas of $p$ in $c$ are paired  (Fig.~\ref{fig:create}, second row, left). Then after acting with $F_p$ both Majoranas will be unpaired  (Fig.~\ref{fig:create}, second row, right).  ${\cal D}(c) + {\cal D}(c_p)$ is an open string containing $2n$ Majoranas, which we label consecutively along the string so that $1$ and $2n$ are the boundary Majoranas.  We let
\beq X_{p,c}=N_{p,c} P_{23} P_{45} \ldots P_{2i, 2i+1} \ldots P_{2n-2,2n-1}
\label{Xbound1} \eeq
with $N_{p,c} = 2^{n/2  -1/2}$. For $|\psi\rangle \in {\cal V}^c$, $F_p |\psi\rangle$ has the same norm as $|\psi\rangle$. Further, $i s_{1,2n} \gamma_{1} \gamma_{2n} F_p |\psi\rangle =  F_p |\psi \rangle$, where $s_{1,2n}$ corresponds to the orientation of the boundary edge $(1,2n)$.

\item Both boundary Majoranas of $p$ in $c$ are unpaired, Fig.~\ref{fig:create}, third row, left. Then after acting with $F_p$ both Majoranas will be paired, Fig.~\ref{fig:create}, third row, right. Again, ${\cal D}(c) + {\cal D}(c_p)$ is an open string with $2n$ Majoranas, $1$ and $2n$ being the boundary Majoranas. We let
\beq X_{p,c}=N_{p,c}  P_{12} P_{34} \ldots P_{2i-1,2i} \ldots P_{2n-1,2n} 
\label{Xbound2} \eeq
with $N_{p,c} = 2^{n/2  -1/2}$. Now, $F_p |\psi\rangle$ has the same norm as $|\psi\rangle$ if $i s_{1,2n} \gamma_1 \gamma_{2n} |\psi \rangle = |\psi\rangle$. If $i s_{1,2n} \gamma_1 \gamma_{2n} |\psi \rangle = - |\psi\rangle$ then $F_p |\psi \rangle  =0$. 

\item One boundary Majorana  of $p$ in $c$ is paired and the other is unpaired, Fig.~\ref{fig:create}, fourth row, left. Then ${\cal D}(c) + {\cal D}(c_p)$ is an open string containing $2n-1$ Majoranas, which we label consecutively. We let $1$ be the initially unpaired Majorana and $2n- 1$ be the initially paired Majorana. After the flip, $1$ is paired and $2n-1$ is unpaired, Fig.~\ref{fig:create}, fourth row, right. Then
\beq X_{p,c}=N_{p,c}  P_{12} P_{34} \ldots P_{2i-1,2i} \ldots P_{2n-3,2n-2} \label{Xbound3} \eeq
with $N_{p,c} = 2^{n/2-1/2}$. Again, $F_{p} |\psi\rangle$ has the same norm as $|\psi\rangle$. 

The derivation of the above properties is sketched in Fig.~\ref{fig:create}.

\end{itemize}

\subsection{Properties of plaquette flip operators}
\label{sec:Fp}

We now list some useful properties of plaquette flip operators $F_p$.

\begin{prop}
\label{prop:allcommute}
The bulk \(F_p\)'s commute with one another, and the boundary \(F_p\)'s all commute with all the bulk \(F_p\)'s. Nearest neighbor boundary \(F_p\)'s do not commute, but otherwise boundary \(F_p\)'s do. 
\end{prop}
We present the proof of this result in appendix \ref{app:Commbulk}. A key consequence of this result is that for a boundary plaquette $p$, $F_p$  maps the ground state manifold of $H_{bulk}$, (\ref{Hbulk2}), associated with the boundary degeneracy, into itself. 

We now focus on the properties of boundary plaquette operators. Let us focus on one component of the boundary at a time; each component is a circle. Going clockwise around the boundary (with the bulk to the right and the vacuum to the left), label the plaquettes as $i = 1, 2, \ldots N$. Label the boundary Majorana shared by  plaquettes $i$ and $i+1$ as $\gamma_{i, i+1}$, see Fig.~\ref{fig:Rbound}. Let $s_i = \pm 1$ be the orientation of the boundary edge from $\gamma_{i-1,i}$ to $\gamma_{i,i+1}$.

\begin{prop}
\label{prop:square}
Let $i$ be a boundary plaquette, and $\gamma_{i-1,i}$, $\gamma_{i, i+1}$ - its boundary Majoranas. If both $\gamma_{i-1,i}$ and $\gamma_{i,i+1}$ are unpaired
in $|\psi\rangle$, $F^2_i |\psi\rangle = \frac12 (1 + i s_{i} \gamma_{i-1,i} \gamma_{i,i+1}) |\psi\rangle$. Otherwise, $F^2_i |\psi\rangle = |\psi\rangle$. 
\end{prop}

\begin{prop}
\label{prop:nncomm}
Consider adjacent boundary plaquettes $i$, $i+1$, and  a state $|\psi\rangle$ where the boundary Majorana $\gamma_{i,i+1}$ shared by these plaquettes is paired (i.e. plaquettes $i$ and $i+1$ have the same spin). Then,
\begin{itemize} 
\item If the other boundary Majorana  $\gamma_{i-1,i}$ on plaquette $i$ is paired,  and the other boundary Majorana $\gamma_{i+1,i+2}$ on plaquette $i+1$ is paired, $[F_i, F_{i+1}] |\psi \rangle = 0$. 
\item If the other boundary Majorana  $\gamma_{i-1,i}$ on plaquette $i$ is unpaired,  and the other boundary Majorana $\gamma_{i+1,i+2}$ on plaquette $i+1$ is unpaired, $[F_i, F_{i+1}] |\psi \rangle = 0$.
\item If the other boundary Majorana $\gamma_{i-1,i}$ on plaquette $i$ is unpaired, and the other boundary Majorana $\gamma_{i+1,i+2}$ on plaquette $i+1$ is paired, $F_{i} F_{i+1} |\psi \rangle = \frac{1}{\sqrt{2}} F_{i+1} F_i |\psi\rangle$. 
\item If the other boundary Majorana $\gamma_{i-1,i}$ on plaquette $i$ is paired, and the other boundary Majorana $\gamma_{i+1,i+2}$ on plaquette $i+1$ is unpaired, $F_{i+1} F_{i} |\psi \rangle = \frac{1}{\sqrt{2}} F_{i} F_{i+1} |\psi\rangle$. 

\end{itemize}
\end{prop}

\begin{prop}
\label{prop:fermtrans}
Let \(i\) be a boundary plaquette, and  $|\psi\rangle$ a state where \(\gamma_{i-1,i}\) is unpaired but \(\gamma_{i,i+1}\) is paired. Then,  \(F_i\gamma_{i-1,i} | \psi\rangle= s_{i} \gamma_{i,i+1} F_i  |\psi\rangle \). 
Likewise, if \(\gamma_{i-1,i}\) is paired but \(\gamma_{i,i+1}\) is unpaired then $F_i \gamma_{i,i+1} |\psi\rangle = s_i \gamma_{i-1,i} F_i |\psi\rangle$.
\end{prop}

The proofs of propositions \ref{prop:square}, \ref{prop:fermtrans} are elementary, while the proof of proposition \ref{prop:nncomm} is given in appendix \ref{app:Commbound}.

\subsection{Boundary Hilbert Space}
We now introduce our labelling of the boundary Hilbert space. To specify a state, we must give the values of the boundary spins and the fusion state of the unpaired Majoranas. Let us focus on one boundary component at a time, labelling the boundary plaquettes and Majoranas as in section \ref{sec:Fp}. We begin with a state $|+\rangle$ where all the boundary spins are $``+"$ and so, there are no unpaired Majoranas. Consider creating and growing a string of $``-"$ spins stretching from plaquette $i$ to plaquette $j$:
\beq |\psi\rangle = F_{(i,j)}  |+\rangle, \quad F_{(i,j)} = F_{j} F_{j-1} \ldots F_{i+2} F_{i+1} F_i \label{Fij} \eeq
According to the discussion below Eqs.~(\ref{Xbound1}), (\ref{Xbound3}) and proposition \ref{prop:fermtrans}, $|\psi\rangle$ is a normalized state with two unpaired Majoranas $\gamma_{i-1,i}$ and $\gamma_{j,j+1}$,  which are fused as
\beq i s_{(i,j)} \gamma_{i-1,i} \gamma_{j,j+1} |\psi\rangle = |\psi\rangle \eeq
where we define 
\beq s_{(i,j)} = s_i s_{i+1} \ldots s_{j-1} s_j \eeq
The state with the opposite fusion channel of the Majoranas is then $\gamma_{i-1,i} |\psi\rangle$. 

We can generalize the above discussion to specify an arbitrary boundary state. We do so by giving the location of the $``-"$ domains and the fusion state of the boundary Majoranas on each  $``-"$ domain. Suppose we have $N_d$ $``-"$ domains labelled by $l = 1 \ldots N_d$ with the $l$'th domain occupying plaquettes $i_l, i_l + 1,i_l + 2 \ldots j_l-1, j_l$, see Fig.~\ref{fig:U11}, top. Note that operators $F_{(i_l, j_l)}$, Eq.~(\ref{Fij}), with different $l$ (corresponding to different $``-"$ domains) commute. 
Then
\beq |\psi\rangle = \prod_{l =1}^{N_d} F_{(i_l, j_l)} |+\rangle  \label{psimultd} \eeq
is a normalized state with the right values of the boundary spins, and with the boundary Majoranas on the $``-"$ domains  fused as,
\beq i s_{(i_l, j_l)} \gamma_{i_l-1,i_l} \gamma_{j_l,j_l+1}  |\psi\rangle  = |\psi\rangle\eeq
 To access the full set of $2^{N_d}$ states with Majoranas fused as
\beq i s_{(i_l, j_l)} \gamma_{i_l-1,i_l} \gamma_{j_l,j_l+1}   \sim  (-1)^{\lambda_l}, \quad \lambda_l  =0,1 \label{fuslambda}\eeq
we must then act on $|\psi\rangle$ in (\ref{psimultd}) with $\gamma_{i_1-1, i_1}^{\lambda_1} \gamma_{i_2-1, i_2 }^{\lambda_2} \gamma_{i_3-1, i_3}^{\lambda_3} \ldots \gamma_{i_{N_d}-1, i_{N_d}}^{\lambda_{N_d}}$. (We choose a convention, where we act with $\gamma$'s on the left boundary of each $``-"$ domain - we could have likewise acted with the $\gamma$'s on the right boundary.) The phase of the  above product  depends on the  order of $\gamma$'s. We, thus, introduce a ``basepoint"  located at the boundary of plaquettes $B-1$ and $B$. It will be useful to leave the basepoint arbitrary. We then define a state,

\bea && |\{i_l, j_l, \lambda_l\}^-; (B-1, B) \rangle = \gamma^{n_{B-1,B}}_{B-1, B} (s_{(B, B)} \gamma_{B,B+1})^{n_{B,B+1}} (s_{(B, B+1)}  \gamma_{B+1,B+2})^{n_{B+1,B+2}}  \nn\\
&& \quad \times 
\ldots (s_{B,k} \gamma_{k, k+1})^{n_{k,k+1}} \ldots  (s_{(B, B-3)} \gamma_{B-3, B-2})^{n_{B-3,B-2}} (s_{(B, B-2)} \gamma_{B-2, B-1})^{n_{B-2,B-1}}  \nn\\
&& \quad \times \prod_{l = 1}^{N_d} F_{(i_l, j_l)} |+\rangle  \label{nstring} \eea
Here, $n_{k,k+1}= \lambda_l$ if $k+1 = i_l$, and $n_{k,k+1} = 0$ otherwise. The prefactors $s_{B,k}$ in front of $\gamma_{k,k+1}$ are inserted for future convenience. 
The superscript $``-"$ on $\{i_l, j_l, \lambda_l\}$ reminds us that the location of the ``$-$" domains is given; the label $l$ runs over $l  =1 \ldots N_d$. The  state (\ref{nstring}) satisfies Eq.~(\ref{fuslambda}). The states  (\ref{nstring}) form an orthonormal basis that spans the boundary Hilbert space.

It is useful to study the dependance of the state $|\{i_l, j_l, \lambda_l\}^-; (B-1, B) \rangle$ on the basepoint $B$. The dependence comes entirely from the Majorana string in (\ref{nstring}) that acts on $\prod_{l = 1}^{N_d} F_{(i_l, j_l)} |+\rangle$. We have,
\beq |\{i_l, j_l, \lambda_l\}^-; (B, B+1) \rangle = ((-1)^{\cal F} \eta)^{n_{B-1,B}} (s_B)^{\cal F} |\{i_l, j_l, \lambda_l\}^-; (B-1, B) \rangle \label{Bshiftlambda}\eeq
where, 
\beq {\cal F} = \sum_{l=1}^{N_d} \lambda_l \,\, (mod\,\,2) \label{Flambda} \eeq
 is the fermion parity  and 
\beq \eta = -s_1 s_2 \ldots s_N = \left\{\begin{array}{rl} 1, & \quad {\rm NS}\\-1, &\quad {\rm R}\end{array}\right.\eeq
with NS and R standing for the Neveu-Schwarz and Ramond spin-structures around the boundary. 

As explained in section \ref{sec:model}, there exists a more convenient labelling for the states (\ref{nstring}) in which locality is more manifest. Namely, given $\{i_l, j_l, \lambda_l\}$, we assign to every plaquette $j = 1 \ldots N$, a label $a_j \in \{0,\frac12,1\} \sim \{1,\sigma,f\}$. More precisely, every $``-"$ plaquette is assigned $a_j = \frac12 \sim \sigma$,  and  every plaquette of a given  $``+"$ domain is assigned the same label $a_j \in \{0,1\} \sim \{1,f\}$, such that the labels $a_{i_l-1}$  and $a_{j_l + 1}$ on the $``+"$ domains neighbouring the $l$'th $``-"$ domain satisfy 
\beq a_{i_l-1} + a_{j_l + 1} = \lambda_l\,\,(mod\,\,2) \label{aalambda} \eeq
 Strictly speaking, such an assignment is only possible if the  fermion parity (\ref{Flambda}), ${\cal F} = 0 \,\,(mod\,\,2)$. Let's focus on this case for now. There is also a difficulty that for every $\{\lambda_l\}$ there are two assignments, $a_j$ and $a'_j$, satisfying (\ref{aalambda}), related by $a'_j = 
\bar{a}_j = 1-a_j$. We utilize both assignments and define:
\begin{itemize}
\item $(-1)^{\cal F} = 1, {\rm NS}$
\beq |\{i_l, j_l, \lambda_l\}^-; (B-1, B) \rangle = \frac{1}{\sqrt{2}}(|\{a_j\};(B-1,B); S_\Phi = 1\rangle +  |\{\bar{a}_j\};(B-1,B); S_\Phi = 1\rangle)\label{adefNS} \eeq
\item $(-1)^{\cal F} = 1, {\rm R}$
\beq |\{i_l, j_l, \lambda_l\}^-; (B-1, B) \rangle = \frac{u}{\sqrt{2}}(|\{a_j\};(B-1,B); S_\Phi = 1\rangle -  |\{\bar{a}_j\};(B-1,B); S_\Phi = 1\rangle)\label{adefR}\eeq
where $u = \pm 1$ depending on the value of spins on plaquettes $B-1$ and $B$:
\bea (\tau^z_{B-1}, \tau^z_B)&\neq& (+,-): \quad u = (-1)^{{a}_R} \nn\\
 (\tau^z_{B-1}, \tau^z_B) &=& (+,-): \quad u = (-1)^{{a}_L} \label{udef}\eea
with $a_R = 0,1$ -- the value of $a$ on the first $``+"$ domain to the right of the $(B-1,B)$ cut, and $a_L = 0,1$ -- the value of $a$ on the first $``+"$ domain to the left of the $(B-1, B)$ cut (i.e. $a_L  = a_{B-1}$). 
\end{itemize}
We will discuss the  meaning of the label $S_\Phi  = 1$ shortly. Note that Eq.~(\ref{adefR}) is well-defined (i.e. invariant under replacing $a \to \bar{a}$). As we will discuss below, the reason for the elaborate choice of the phase factor $u$ in Eq.~(\ref{adefR}) is that the resulting state $|\{a_j\}; (B-1,B)\rangle$ has simple transformation properties under the change of the basepoint $B$. 

We now extend the above discussion to the case of odd fermion parity ${\cal F} = 1 \,\,(mod \,\,2)$. As in section \ref{sec:Hilbert}, we let the labels $\{a_j\}$ live on the double cover of the boundary circle with sites $j = 1\ldots 2N$. We then extend the spins $\tau^z_j$ periodically, so that $\tau^z_{j+N} = \tau^z_j$ and we have $2N_d$ domains, $l = 1\ldots 2N_d$, on the double-cover. We also extend $\lambda_l$ periodically so that $\lambda_{l +  N_d}  = \lambda_l$. We then solve Eq.~(\ref{aalambda}) for $a_j$, $j =1\ldots 2N$. Again, there are two solutions $a_j$ and $a'_j$ related by $a'_j = \bar{a}_j$. Further, each solution satisfies, $a_{j+N} = \bar{a}_j$. We now define,
\begin{itemize}
\item $(-1)^{\cal F} = -1, {\rm NS}$
\beq |\{i_l, j_l, \lambda_l\}^-; (B-1, B) \rangle = \frac{u}{\sqrt{2}}(|\{a_j\};(B-1,B); S_\Phi = -1\rangle -  |\{\bar{a}_j\};(B-1,B); S_\Phi  =-1\rangle)\label{adefNSodd}\ \eeq
\item $(-1)^{\cal F} = -1, {\rm R}$
\beq |\{i_l, j_l, \lambda_l\}^-; (B-1, B) \rangle = \frac{1}{\sqrt{2}}(|\{a_j\};(B-1,B); S_\Phi = -1\rangle +  |\{\bar{a}_j\};(B-1,B); S_\Phi = -1\rangle) \label{adefRodd}\eeq
with $u = \pm 1$ in Eq.~(\ref{adefNSodd}) still given by Eq.~(\ref{udef}). Note that the base-point $B$ must now be specified modulo $2N$ rather than modulo $N$. 
\end{itemize}

As we discussed in section \ref{sec:model}, we may think of $\{a_j\}\big|_{j=1}^{2N}$  with $a_{j+N} = \bar{a}_j$ as their being a flux of the symmetry $S$, Eq.~(\ref{S}), through the circle. This explains the label $S_\Phi = -1$ in Eqs.~(\ref{adefNSodd}), (\ref{adefRodd}). Likewise, we may also work on the double-cover in the even fermion parity case (\ref{adefNS}), (\ref{adefR}) extending $a_j$ via $a_{j+N} = a_j$. In this case, there is no $S$-flux through the circle, so we label the states as $S_\Phi = +1$. We note that depending on the NS/R spin-structure, the states also carry a definite $S$-charge in accordance with Fig.~\ref{fig:NSR}. Thus, we see that we obtain a bosonized labelling of the Hilbert space which exactly agrees with the discussion in section \ref{sec:Hilbert}. 

Before we continue, let us elaborate on the bosonized labelling of the state $|-\rangle$ where all spins are $``-"$, which is not covered by the definitions above. In the NS sector, this state has the same fermion parity $(-1)^{\cal F}  = 1$ as the $|+\rangle$ state, and we obtain it by acting with a string of plaquette flip operators $F_j$ running around the circle:
\bea  &&(-1)^{\cal F}  =1,\,\, {\rm NS}:\,\, \nn\\
&& |-\rangle_{NS} =  F_{B-1} F_{B-2} \ldots F_{B+2} F_{B+1} F_B |+\rangle_{NS} = |\tfrac12 \tfrac12 \cdots \tfrac12; (B-1, B); S_\Phi = 1\rangle \label{minusNS} \eea
i.e. this is a state with all $a_j  = \frac12$. On the other hand, in the R sector, the state $F_{B-1} F_{B-2} \ldots F_{B+2} F_{B+1} F_B |+\rangle_{R}$ vanishes. In fact, in the R sector, the state $|-\rangle$ has opposite fermion parity to the $|+\rangle$ state. We define,
 \bea &&(-1)^{\cal F}  =-1,\,\, {\rm R}:\,\, \nn\\
&&  |-\rangle_R = Z_{B} |+\rangle_R = |\tfrac12 \tfrac12 \cdots \tfrac12; (B-1, B); S_\Phi = -1\rangle\nn\\
&& Z_B = F_{B-1} \gamma_{B-1,B} F_{B-2} \ldots F_{B+2} F_{B+1} F_B
\label{minusR} \eea

It is useful to consider the transformations of states in our bosonized labelling under a change of base-point $B$. From (\ref{Bshiftlambda}), we find particularly simple transformation properties:
\bea |a; (B, B+1); S_\Phi = 1\rangle &=&  |a; (B-1, B); S_\Phi = 1\rangle \nn\\
 |a; (B, B+1); S_\Phi = -1\rangle &=&  s_B |a; (B-1, B); S_\Phi = -1\rangle \label{Bchange}\eea

\subsection{Operator action}
We now discuss the action of simple boundary operators in the bosonized notation. We first discuss the plaquette flip operator $F_i$ - it acts the same way as $H_i$ in Eq.~(\ref{H3}) does, with $c_1 = c_2=c_3=c_4=c_5 = c_6 =  c_7 = c_8 = 1$. We won't give the full proof here, but in appendix \ref{app:Fbos} we illustrate the proof strategy by discussing the last case ($c_8$) in Eq.~(\ref{H3}).

We also discuss the action of Majorana operators $\gamma_{i,i+1}$ in the bosonized notation. Let us define,
\beq \Gamma^+_{i,i+1} = s_{(B, i)} \gamma_{i,i+1}\frac{1-\tau^z_i \tau^z_{i+1}}{2} \label{Gammap} \eeq
This way, $\Gamma^+_{i,i+1}$ acts within the ground state subspace of $H_{bulk}$. Note that the definition of $\Gamma^+_{i,i+1}$ is with respect to the base-point $B$. We find that in the bosonized notation, the operator (\ref{Gammap}) has the same action as $\Gamma^+_{i,i+1}$ in Eq.~(\ref{Gammapm}). We illustrate part of the proof in appendix \ref{app:Gammabos}.

\subsection{Symmetry action}
We now discuss the action of the $Z_2$ symmetry $U$, Eq.~(\ref{UTF}), of the TF model on the bosonized boundary Hilbert space. A key observation is that 
\beq U F_i U^{\dagger} = F_i \label{UF}\eeq
 Indeed, the location of the Majorana string entering $F_i$ depends only on the domain wall structure; further, $U \gamma_i U^{\dagger}  =\gamma_i$. 

\subsubsection{NS boundary conditions}

We begin with a disk bulk geometry, so that the boundary is a circle. The Kasteleyn condition then means that the boundary has the NS spin structure. Consider the boundary state $|+\rangle$. Under the $Z_2$ action it turns into the state $|-\rangle$, Eq.~(\ref{minusNS}),
\beq U |+\rangle_{NS} = \xi |-\rangle_{NS} = \xi F_{B-1} F_{B-2} \ldots F_{B+2} F_{B+1} F_B |+\rangle_{NS} = \xi |\tfrac12 \tfrac12 \cdots \tfrac12; (B-1, B); S_\Phi = 1\rangle  \label{Up}\eeq
where $\xi$ is a phase. We have $U^2 = 1$, moreover, 
\bea U^2 |+\rangle_{NS} &=& \xi U F_{B-1} F_{B-2} \ldots F_{B+2} F_{B+1} F_B |+\rangle_{NS} = \xi  F_{B-1} F_{B-2} \ldots F_{B+2} F_{B+1} F_B U |+\rangle_{NS} \nn\\
&=& \xi^2 F_{B-1} F_{B-2} \ldots F_{B+2} F_{B+1} F_B |\tfrac12 \tfrac12 \cdots \tfrac12; (B-1, B); S_\Phi = 1\rangle\eea
where we've applied Eq.~(\ref{Up}) in the first step, used Eq.~(\ref{UF}), and then applied (\ref{Up}) again. We now successively apply the rules in Eq.~(\ref{H3}) to find
\beq U^2 |+\rangle_{NS} = \frac{\xi^2}{\sqrt{2}}\left( |00 \cdots 0; (B-1, B); S_\Phi = 1\rangle  + |11 \cdots 1; (B-1, B); S_\Phi = 1 \rangle\right) = \xi^2 |+\rangle_{NS} \eeq
Thus, $\xi = \pm 1$. We won't attempt to fix this sign and will simply use $\xi = 1$. 
Next, we discuss the action of $U$ on an arbitrary state $|\psi\rangle$ in the NS sector. For simplicity, let's take $|\psi\rangle$ to have $(-1)^{\cal F} =1$;  one can derive the symmetry action, Eq.~(\ref{U22}), for the case $(-1)^{\cal F} = -1$   in a similar manner. Let
\beq |\psi\rangle = \frac{1}{\sqrt{2}} (|a; (B-1,B); S_{\Phi} = 1 \rangle + |\bar{a},(B-1,B); S_{\Phi} = 1 \rangle)\eeq
 Let's take the $``-"$ domains in $|\psi\rangle$ to be $(i_l, j_l)$, $l = 1\ldots N_d$, going clockwise consecutively around the circle. Further, let $a_{j_l+1} = \mu_l \in \{0,1\}$, see figure \ref{fig:U11}, top.  We extend $\mu_l$ periodically such that $\mu_{l+N_d} = \mu_l$. Since $U$ does not change the fermion parity in the NS sector, we may choose the basepoint $B$ arbitrarily. Here, we use $B = i_1$. Then
 \beq |\psi\rangle =   \prod_{l=1}^{N_d} (s_{(i_1, i_l-1)} \gamma_{i_l-1,i_l})^{\mu_{l-1} +\mu_{l}} \prod_{l = 1}^{N_d} F_{(i_l, j_l)} |+\rangle_{NS}\eeq
The terms in the first product are arranged with smaller $l$ to the left, and in the $l = 1$ term $s_{i_1, i_1 -1} \equiv  1$.  Now, using Eq.~(\ref{Up}),
\beq U|\psi\rangle = \prod_{l=1}^{N_d} (s_{(i_1, i_l-1)} \gamma_{i_l-1,i_l})^{\mu_{l-1} +\mu_{l}} \prod_{l = 1}^{N_d} F_{(i_l, j_l)} |\tfrac12 \tfrac12 \cdots \tfrac12; (i_1-1, i_1); S_\Phi = 1\rangle \label{Upsi1} \eeq
We now apply the rules (\ref{H3}), (\ref{Gammapm}) to evaluate the above expression. We have:
\beq \prod_{l = 1}^{N_d} F_{(i_l, j_l)} |\tfrac12 \tfrac12 \cdots \tfrac12; (i_1-1, i_1); S_\Phi = 1\rangle = \frac{1}{2^{N_d/2}}\sum_{\{\rho_l\}} |\{ i_l, j_l,\rho_l\}^+; (i_1-1, i_1); S_\Phi = 1\rangle \eeq 
Here, $\rho_l \in \{0,1\}$, $l = 1 \ldots N_d$, and $ |\{ i_l, j_l,\rho_l\}^+; (i_1-1, i_1); S_\Phi = 1\rangle$ denotes a state in the bosonized notation where consecutive  $``+"$ domains stretch from $i_l$ to $j_l$ and carry the label $a = \rho_l$ (like figure \ref{fig:U11}, bottom, but with $\nu_l \to \rho_l$). Next, acting with the string of Majorana operators in (\ref{Upsi1}) using (\ref{Gammapm})
\beq U|\psi\rangle = \frac{1}{2^{N_d/2}} \sum_{\{\rho_l\}} \left(\prod_{l=1}^{N_d} ((-i) (-1)^{\rho_l})^{[\mu_{l-1}+\mu_l]_2} \right) |\{ i_l, j_l,[\rho_l + \mu_l + \mu_{N_d}]_2\}^+; (i_1-1, i_1); S_\Phi = 1\rangle \label{Upsi2} \eeq
Making a change of variables $\nu_l = [\rho_l + \mu_l + \mu_{N_d}]_2$ and simplifying the product in brackets in Eq.~(\ref{Upsi2}),
\beq U|\psi\rangle = \frac{1}{2^{N_d/2}} \sum_{\{\nu_l\}} (-1)^{\sum_{l=1}^{N_d} \mu_l(\nu_l+\nu_{l+1})} |\{ i_l, j_l,\nu_l \}^+; (i_1-1, i_1); S_\Phi = 1\rangle \label{Upsi3} \eeq
which agrees with Eq.~(\ref{U11}).

\subsubsection{R boundary conditions}
We now consider the case of Ramond boundary conditions. We take the bulk to be an annulus. We order plaquettes on the outside boundary clockwise and the plackets on the inside boundary counter-clockwise (such that the bulk is always to the right when going around the boundary). We pick basepoints $B$ and $B'$ on the outside and inside boundaries. Let's start with the $|++\rangle$ boundary state, where the first and second entries refer to outer and inner boundaries respectively.  $U$ maps this to the $|--\rangle$ boundary state,
\beq U|++\rangle_{R} = \xi_{B,B'} Z^{out}_B Z^{in}_{B'} |++\rangle_R \label{UppR}\eeq
with operators $Z^{out}_B$, $Z^{in}_{B'}$ on the outer and inner boundaries of the annulus defined in Eq.~(\ref{minusR}). $\xi_{B,B'}$ is a base-point dependent phase. Indeed, from (\ref{Bchange}), $\xi_{B+1, B'} = s^{out}_B \xi_{B,B'}$ and $\xi_{B, B'+1} = s^{in}_{B'} \xi_{B, B'}$. Further, by utilizing $U^2 = 1$, we find $\xi_{B, B'} = \pm 1$. We guess that $\xi_{B,B'}$ is related to the product of $s_{ij}$ along a path through the bulk of the annulus connecting the two base-points $B$ and $B'$, but we won't attempt to prove this. 

Now, for a general boundary state $|\psi\rangle$, the action of $U$ factorizes as
\beq U|\psi\rangle = i \xi_{B,B'} U^{out}_B U^{in}_{B'} (-1)^{\cal F} |\psi\rangle \label{URfact}\eeq
Here $U^{out}_B$ and $U^{in}_{B'}$ are operators acting on outer and inner boundaries of the annulus, respectively, moreover, both are fermion parity odd: $\{(-1)^{\cal F}, U^{out}_B\} = \{(-1)^{\cal F}, U^{in}_{B'}\} = 0$. Here, $(-1)^{\cal F}$ is the total fermion parity of the system. We could have factorized $(-1)^{\cal F}$ into contributions from outer and inner boundaries and included these in $U^{out}$, $U^{in}$, but we find the above form more convenient. The phase factor $i \xi_{B,B'}$ is also included for convenience. Clearly, $U^{out}$ ($U^{in}$) must be a symmetry of the  Hamiltonian for the outer (inner) boundary.  

Comparing (\ref{UppR}) and (\ref{URfact}), we set
\beq U^{out}_B|+\rangle_{out} = e^{-\pi i/4} Z^{out}_{B} |+\rangle_{out}, \quad U^{in}_{B'} |+\rangle_{in} = e^{-\pi i/4} Z^{in}_{B'} |+\rangle_{in} \label{Uplusplus}\eeq
where $|+\rangle_{out}$ ($|+\rangle_{in}\rangle$) refers to any state with $``+"$ on the outer (inner) boundary.

Next, let us concentrate on the outer boundary and consider states $|\psi\rangle$ where the inner boundary of the annulus is uniformly $``+"$, but the outer boundary is arbitrary. We will use the notation (\ref{adefR}), (\ref{adefRodd}) for these states - i.e. we use  $|++\rangle_R$ instead of $|+\rangle$ on the RHS of  Eq.~(\ref{nstring}). We let the $``-"$ domains in $|\psi\rangle$ be $(i_l, j_l)$, $l = 1 \ldots N_d$. Let us assume $(-1)^{\cal F} = 1$ on $|\psi\rangle$, the case $(-1)^{\cal F} = -1$ can be treated similarly. Let 
\beq |\psi\rangle = \frac{1}{\sqrt{2}} \left(|a; (B, B-1); S_{\Phi} = 1\rangle - |a; (B, B-1); S_{\Phi} = 1\rangle\right) \eeq
Since $U^{out}_B$ changes the fermion parity of the state, by Eq.~(\ref{Bchange}), the $B$ dependence of $U^{out}_B$ cancels with the $B$ dependence of $\xi_{B,B'}$ in Eq.~(\ref{URfact}). Therefore, we may choose $B$ freely. Let us pick $B  =i_1$. Further, label $\mu_l = a_{j_l+1}$, $l = 1 \ldots N_d$, and extend $\mu_{l}$ so that $\mu_{l+ N_d} = \mu_l$. Then by Eq.~(\ref{adefR}),
 \bea |\psi\rangle =  A^{out} |++\rangle_{R}, \quad\quad A^{out} =  (-1)^{\mu_{N_d}} \prod_{l=1}^{N_d} (s^{out}_{(i_1, i_l-1)} \gamma^{out}_{i_l-1,i_l})^{\mu_{l-1} +\mu_{l}} \prod_{l = 1}^{N_d} F^{out}_{(i_l, j_l)}\label{psiUR}\eea
As before, the terms in the first product are arranged with smaller $l$ to the left, and in the $l = 1$ term $s^{out}_{i_1, i_1 -1} \equiv  1$. The superscripts $out$ remind us that the operators act on the outer boundary. Acting with $U$ on $|\psi\rangle$ above, and using Eqs.~(\ref{URfact}), (\ref{UppR}),
\beq i U^{out}_B U^{in}_{B'} |\psi\rangle = A^{out} Z^{out}_B Z^{in}_{B'} |++\rangle_R\eeq 
Using (\ref{Uplusplus}), $U^{in}_{B'}|\psi\rangle = e^{-\pi i/4} Z^{in}_{B'} |\psi\rangle$. Further, using $(Z^{in}_{B'})^2|+\rangle_{in} = i |+\rangle_{in}$ and $\{Z_{B'}, U^{out}_B\} = \{Z_{B'}, Z^{out}_B\} = [Z_{B'}, A^{out}] = 0$,
\beq U^{out}_{B}|\psi\rangle = e^{-\pi i/4} A^{out} Z^{out}_B |++\rangle_R  =e^{-\pi i/4} A^{out} |\tfrac12 \tfrac12 \cdots \tfrac12; (B-1, B); S_\Phi = -1\rangle  \eeq
We now evaluate the RHS of equation above using (\ref{H3}), (\ref{Gammapm}). We have 
 \beq \prod_{l = 1}^{N_d} F^{out}_{(i_l, j_l)} |\tfrac12 \tfrac12 \cdots \tfrac12; (i_1-1, i_1); S_\Phi = -1\rangle =    \frac{1}{2^{N_d/2}}\sum_{\{\rho_l\}} |\{ i_l, j_l,\rho_l\}^+; (i_1-1, i_1); S_\Phi = -1\rangle  \label{FpsiR}\eeq
Here, $\rho_l \in \{0,1\}$, $l = 1 \ldots N_d$, and $|\{ i_l, j_l,\rho_l\}^+; (i_1-1, i_1); S_\Phi = -1\rangle$ denotes a state with $``+"$ domains $\{i_l, j_l\}$ carrying $a = \rho_l$. Since this state is in the $S_{\Phi} = -1$ sector, it is really defined on the double cover, so we may extend $l  = 1 \ldots 2 N_d$ and $i_{l+ N_d} = i_l+N$, $j_{l+N_d} = j_l + N$, $\rho_{l+N_d} = 1-\rho_l$ (see Fig.~\ref{fig:U21}, bottom, with $\nu_l \to \rho_l$.) Now acting with the Majorana string in $A^{out}$ on (\ref{FpsiR}),
\bea U^{out}_{i_1}|\psi\rangle =  \frac{e^{-\pi i/4} (-1)^{\mu_{N_d}}}{2^{N_d/2}} \sum_{\{\rho_l\}} \left(\prod_{l=1}^{N_d} ((-i) (-1)^{\rho_l})^{[\mu_{l-1}+\mu_l]_2} \right) |\{ i_l, j_l,[\rho_l + \mu_l + \mu_{N_d}]_2\}^+; (i_1-1, i_1); S_\Phi = -1\rangle \nn\\ \label{UpsiR1}\eea
Let's make a change of variables $\nu_l = [\rho_l + \mu_l + \mu_{N_d}]_2$, which is again defined for $l = 1 \ldots 2 N_d$ such that $\nu_{l+N_d} = 1-\nu_l$. Then simplifying the phase in brackets in Eq.~(\ref{UpsiR1}),
\beq U^{out}_{i_1} |\psi\rangle = \frac{e^{-\pi i/4}}{2^{N_d/2}} \sum_{\{\nu_l\}} (i)^{\sum_{l = 1}^{2 N_d} \mu_l [\nu_{l+1}-\nu_l]_2} |\{i_l, j_l, \nu_l\}; (i_1-1,i_1); S_\Phi = -1\rangle \eeq
which exactly agrees with Eq.~(\ref{U21}).

\section{Discussion}
\label{sec:disc}
In this paper we have shown how to mimick the edge of 2d beyond supercohomology fermion SPTs  in a strictly 1d model. This required using a Hilbert space, which is not a local tensor product, but rather is obtained from a local tensor product by imposing a local constraint. While if one ignores the symmetry of the model it is trivial to extend the Hilbert space to a tensor product Hilbert space, we expect that the action of the symmetry cannot be extended. It would be interesting to characterize this obstruction, similar to the algebraic characterization of obstructions to decomposing a finite depth unitary symmetry as an onsite symmetry.\cite{ElseNayak} In fact, we have argued that the edge cannot be mimicked with a local tensor product (fermionic) Hilbert space and a locality preserving unitary symmetry action. Our argument relied on  the classification of locality preserving unitaries in 1d fermion systems.\cite{FloquetF} It would be interesting to understand more precisely how a constrained Hilbert space alters the classification  in Ref.~\onlinecite{FloquetF} and related classifications in bosonic systems.\cite{FloquetC}

We would like to point out that there are other examples where the edge of SPTs can be mimicked by using a constrained Hilbert space.\cite{SonHLL, ChongDual, MVDual, SenthilHLL, flatband} The most well-known of these occurs for 3d fermion SPTs with symmetry $U(1) \times Z^T_2$ (non-interacting class AIII). Here, the anti-unitary time-reversal symmetry ${\cal T}$ commutes with particle number $U(1)$, i.e. ${\cal T} Q {\cal T}^{\dagger} = -Q$, with $Q$ - the $U(1)$ charge. Thus, ${\cal T}$ is really an anti-unitary particle-hole symmetry. In the absence of interactions SPTs with this symmetry are classified by an integer  $n \in Z$, which is reduced to $Z_8$ by interactions.\cite{Wang2014,MCFV2014}\footnote{Interactions also give rise to an entirely new phase, making the full classification $Z_8 \times Z_2$.\cite{FreedHopkins}} The surface of the generating phase $n  =1$ can be mimicked in 2d in the following way. Consider a spinless 2d electron gas in a magnetic field $B$: the system will form Landau levels. Consider the Hilbert space of just the lowest Landau level.\footnote{In fact, any Landau level will do.} One can now define an anti-unitary particle-hole symmetry acting within the lowest Landau level - this precisely mimicks the action of ${\cal T}$ on the surface. Crucially, in the 2d model the action of particle-hole symmetry is only defined in the constrained lowest Landau level Hilbert space. We note that one difference with our construction presented in this paper is that in the Landau level example  the many-body Hilbert space is built out of single-body wave-functions which are constrained. We don't know of a way to decompose the Hilbert space of our 1d model in a similar way and, in fact, the scaling of the Hilbert space dimension with system size, Eq.~(\ref{Vdim}), suggests that it is impossible. We also point out that the $U(1)$ symmetry is not essential for the stability of the $U(1) \times Z^T_2$ SPT phase or for the effective 2d boundary model; one can break $U(1) \times Z^T_2 \to Z^T_4$. The resulting $Z^T_4$ group generated by time-reversal symmetry ${\cal T}$ with  ${\cal T}^2 = (-1)^{\cal F}$ is the same symmetry class (DIII in the non-interacting nomenclature)  that we discussed in section \ref{sec:odds}: in 3d the non-interacting phases are classified by $m \in Z$; this classification is reduced to $Z_{16}$ by interactions.\cite{KitaevZ16, Fidkowski2013, Wang2014, MCFV2014, KapustinFerm, FreedHopkins}  The $n = 1$ phase of $Z_8$ with $U(1)\times Z^T_2$ symmetry becomes the $m  = 2$ phase of $Z_{16}$ with $Z^T_4$ symmetry. It has been shown in Ref.~\onlinecite{WenSuper2018} that only phases with $m = 0 \,\,({\rm mod} \,\, 4)$ are contained in the supercohomology classification. Thus, $m  =2$ is a beyond supercohomology phase.

In the light of the above examples one may wonder whether the boundaries of all beyond supercohomology SPTs can be recreated  in a  constrained Hilbert space, whether such a Hilbert space is a necessary requirement, and whether it suffices for the constraint to be local. We leave these questions for future work.


\vspace{0.3cm}

{\it Note added:} While this work was being finalized, Ref.~\onlinecite{RyanBos} appeared, which studies bosonization of fermion SPT phases in the presence of boundaries. We have also learned about a forthcoming work, Ref.~\onlinecite{LevinF}, whose results partially overlap with those reported here.

 
\acknowledgements
We are grateful to Ryan Thorngren for sharing his work, Ref.~\onlinecite{RyanBos}, on bosonization of fermion SPT boundaries prior to publication. We also thank Kyle Kawagoe and Michael Levin for sharing their forthcoming work, Ref.~\onlinecite{LevinF}, with us. We thank Cecile Repellin for guidance with numerical simulations. We also thank Dominic Else, Tarun Grover, Ying Ran, Nathanan Tantivasadakarn, Ashvin Vishwanath and Liujun Zou for discussions. R.~A.~J. is supported by the National Science Foundation Graduate Research Fellowship under Grant No. 1122374.

\appendix
\section{Commutation relations of $F_p$'s}
\label{app:Comm}

\subsection{At most one boundary plaquette}
\label{app:Commbulk}

We will show that the commutator $[F_p, F_q]|\psi\rangle = 0$ for $|\psi\rangle \in {\cal V}^c$, when at most one of $p$, $q$ is a boundary plaquette.
 First, we expand out the commutator:
\begin{align*}
\left[F_p,F_q\right]&=\left[\sum_cX_{p,c}\otimes(\tau^x_{p}P_{c}),\sum_{c'}X_{q,c'}\otimes(\tau^x_{q}P_{c'})\right]\\
&=\sum_{c,c'}(X_{p,c'}X_{q,c})\otimes(\tau^x_{p}P_{c'}\tau^x_{q}P_{c})-\sum_{c,c'}(X_{q,c'}X_{p,c})\otimes(\tau^x_{q}P_{c'}\tau^x_{p}P_{c})
\end{align*}
where we have relabelled \(c\) and \(c'\) in the first sum. Strictly speaking, $c$ in the first line is summed over all the spin configurations of $p$ and all its neighbouring plaquettes, while $c'$ is summed over the spin configurations of $q$ and all its neighbouring plaquettes, and $P_c$, $P_c'$ denote corresponding projectors. However, we  extend each sum and projector to spin configurations of $p$, $q$ and all their neighbouring plaquettes. Notice, in the first sum in the second line \(P_{c'}\tau_{q}^xP_c\) can only be nonzero if \(c'\) is \(c\) with the spin at \(q\) flipped; call this \(c_q\). Similarly, in the second sum \(P_{c'}\tau_p^xP_c\) can only be nonzero if \(c'\) is \(c\) with the spin at \(p\) flipped; call this \(c_p\). Then, we have:
\begin{align*}
\left[F_p,F_q\right]&=\sum_{c}(X_{p,c_q}X_{q,c})\otimes(\tau^x_{p}\tau^x_{q}P_{c})-(X_{q,c_p}X_{p,c})\otimes(\tau^x_{p}\tau^x_{q}P_c),
\end{align*}
where we have used that the \(\tau\)'s commute. Thus, to show that \(F_p\) and \(F_q\) commute, it suffices to show:
\[X_{p,c_q}X_{q,c} = X_{q,c_p}X_{p,c}\]
acting on a state consistent with \(c\)'s dimer covering. Recall that we defined in Eqs.~(\ref{Xbulk}), (\ref{Xbound1}), (\ref{Xbound2}), (\ref{Xbound3}),  $X_{q,c}=N_{q,c}\Pi_{q,c}$, where $\Pi_{q,c}$ is a product of projectors $P_{ij} = \frac{1}{2}(1+i s_{ij} \gamma_{i}\gamma_{j})$ over dimers $(ij)$ in ${\cal D}_{c_q} \cap ({\cal D}_c+ {\cal D}_{c_q})$ and \(N_{q,c}\) is a positive real normalization. 
Thus, we want to show
\beq N_{p,c_q}N_{q,c}  \Pi_{p,c_q} \Pi_{q,c}  = N_{q,c_p}N_{p,c}  \Pi_{q,c_p}\Pi_{p,c} \label{NPi} \eeq
on a state consistent with $c$ dimer covering. 

\begin{figure}[h!]
\centering
\includegraphics[width=0.4\textwidth]{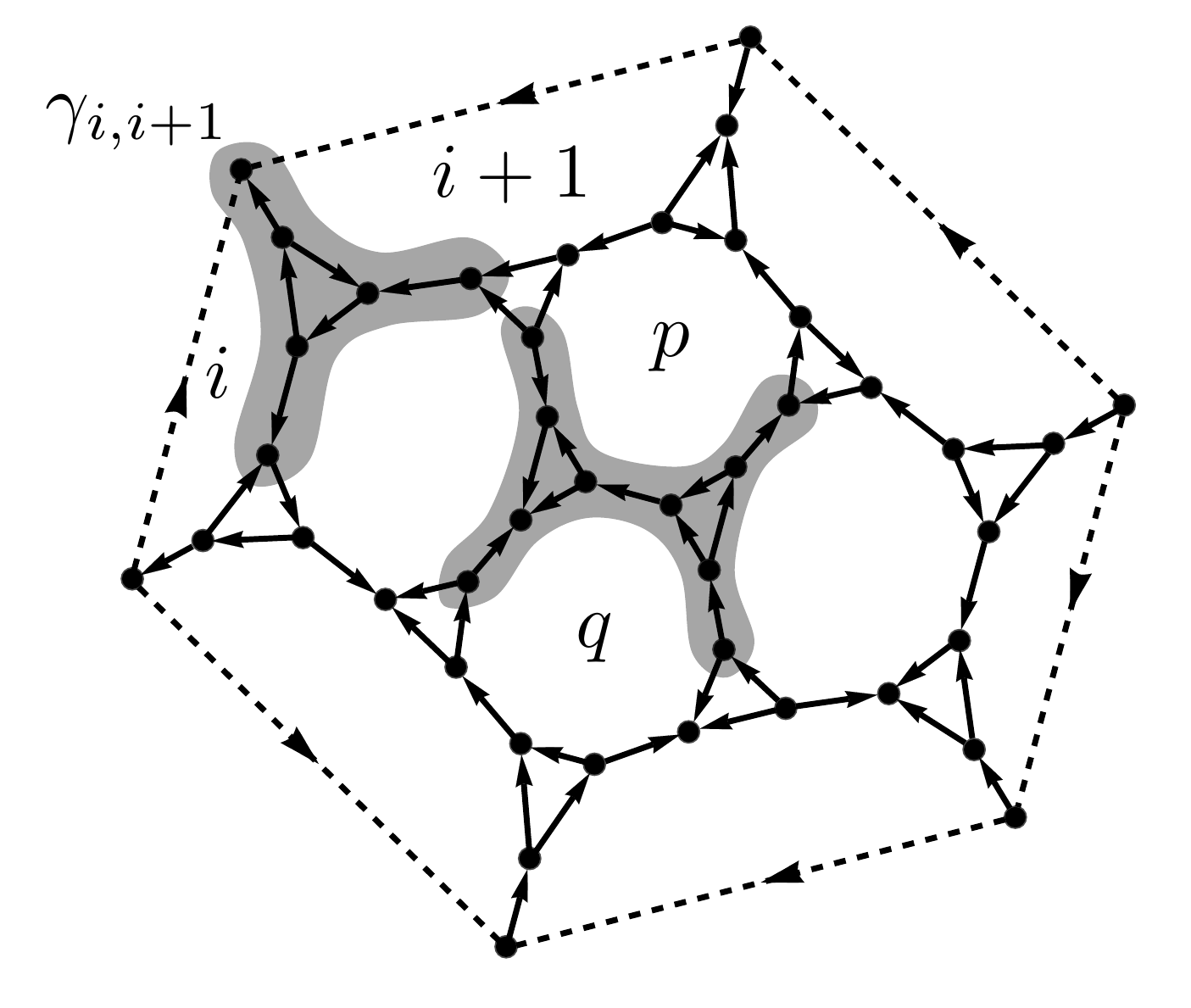}
\caption{Center grey: the region $R_{p,q}$ relevant for studying the commutator $[F_p, F_q]$, where at most one of $p$,$q$ is a boundary plaquette. $R_{p,q}$ consists of 11 edges and 10 Majoranas. 
Left-top grey: the region $R_{i,i+1}$ relevant for studying the commutator $[F_i, F_{i+1}]$ of boundary plaquette flip operators. $R_{i,i+1}$ consists of 6 edges and 6 Majoranas. 
}
\label{fig:Rbound}
\end{figure}

Clearly, if $p$ and $q$ do not share any edges then the equality (\ref{NPi}) holds. So let's consider the case when $p$ and $q$ are adjacent. The interesting behavior will be near the edge shared by \(p\) and \(q\), in the region $R_{p,q}$ shown in the center of Fig.~\ref{fig:Rbound}. This region includes the 11 edges that connect to the 6 Majoranas in the intersection $\d'p \cap \d'q$. Let's decompose each projector in (\ref{NPi}) as $\Pi_{r,\tilde{c}} = \Pi^{out}_{r,\tilde{c}} \Pi^{in}_{r,\tilde{c}}$, $r = p,q$, where $\Pi^{in}$ involves the projectors over edges in $R_{p,q}$ and $\Pi^{out}$  - the projectors over edges outside $R_{p,q}$. We note that the projector strings  in $X_{p,c_q}$ and $X_{p,c}$ coincide outside of $R_{p,q}$ (similarly, for $X_{q,c_p}$ and $X_{q,p}$). Therefore, $\Pi^{out}_{p,c_q} = \Pi^{out}_{p,c}$ and $\Pi^{out}_{q,c_p} = \Pi^{out}_{q,c}$. Further, $[\Pi^{out}_{q,c}, \Pi^{in}_{p,c_q}]=[\Pi^{out}_{p,c}, \Pi^{in}_{q,c_p}] =  [\Pi^{out}_{p,c}, \Pi^{out}_{q,c}] = 0$. Thus, it is enough to prove 
\beq N_{p,c_q}N_{q,c}  \Pi^{in}_{p,c_q} \Pi^{in}_{q,c}  = N_{q,c_p}N_{p,c}  \Pi^{in}_{q,c_p}\Pi^{in}_{p,c} \label{NPi2} \eeq
on a state $|\psi\rangle$ consistent with $c$ dimer covering. Let $\Pi^{in}_c$ be the projector onto the dimers in $c$ that lie in region $R_{p,q}$. Since, $\Pi^{in}_c |\psi\rangle = |\psi\rangle$, it suffices to prove 
\beq N_{p,c_q}N_{q,c}  \Pi^{in}_{p,c_q} \Pi^{in}_{q,c} \Pi^{in}_c = N_{q,c_p}N_{p,c}  \Pi^{in}_{q,c_p}\Pi^{in}_{p,c} \Pi^{in}_c \label{NPi3}\eeq
We further note that the ratios $N_{p,c_q}/N_{p,c}$ and $N_{q,c_p}/N_{q,c}$ can be determined just from $c$ on the four plaquettes in $R_{p,q}$ (again, since the projector strings coincide outside $R_{p,q}$). 

\begin{figure}[h!]
\centering
\includegraphics[width=\textwidth]{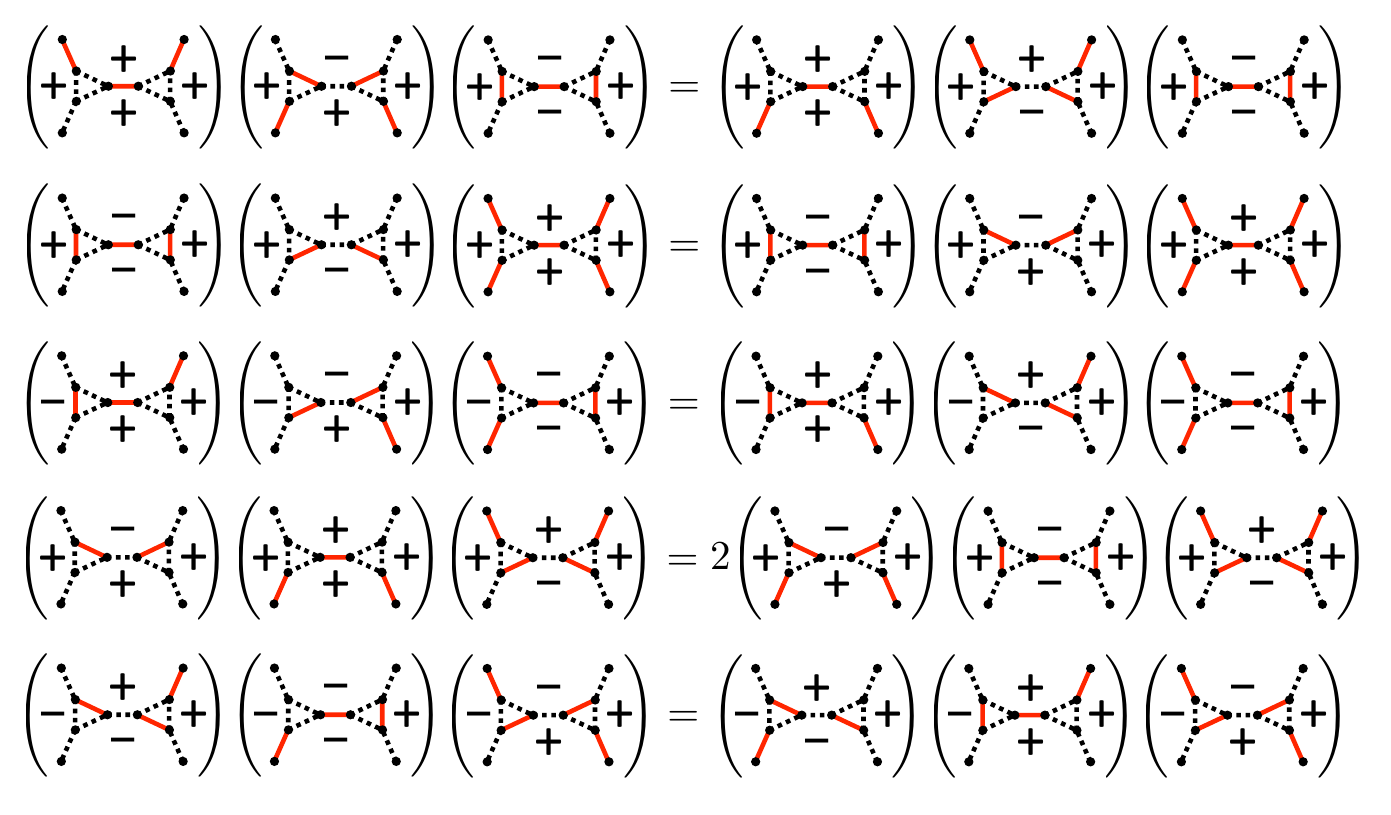}
\caption{ Minimal identities of projectors needed to prove commutativity of plaquette flip operators, Eq.~(\ref{NPi3}). Each diagram in brackets denotes a product of projectors $P_{ij} = \frac12(1+i s_{ij} \gamma_i \gamma_j)$ over the links $(ij)$ shown in red. The top center plaquette is $p$ and the bottom center plaquette is $q$.
}
\label{fig:rules}
\end{figure}

The rest of the proof involves checking Eq.~(\ref{NPi3}) for each of the 16 spin configurations $c$ on the four plaquettes in $R_{p,q}$. By symmetry, we can reduce this to 5 cases, which are shown in Fig.~\ref{fig:rules}. Each diagram in brackets in Fig.~\ref{fig:rules} denotes a product of projectors $P_{ij}$ over the edges marked in solid red, and the operators in brackets are multiplied. The diagrams on the left of each equation in Fig.~\ref{fig:rules} correspond to  $\Pi^{in}_{p,c_q} \Pi^{in}_{q,c} \Pi^{in}_c$ and the diagrams on the right correspond to $\Pi^{in}_{q,c_p}\Pi^{in}_{p,c} \Pi^{in}_c$. 
The identities in Fig.~\ref{fig:rules} can be proved by brute force. (Note that the identity in row 2 is just the hermitian conjugate of the identity in row 1). Notice that in rows 1, 2, 3, and 5, the combined lengths of the paths along which the dimers are shifted on the left and right sides of Eq.~(\ref{NPi}) are the same, so the product of \(N\)'s on the LHS and RHS are equal. On the other hand, in row 4, the combined length of the path is 4 Majoranas longer in the right ordering than the  left ordering, and the resulting factor of 2 difference in the \(N\)'s cancels with the factor of 2 from the projector identity.

We end by noting that the proof above is independent of whether both $p$ and $q$ are bulk plaquettes or one of them is a boundary plaquette.

\subsection{Boundary plaquettes}
\label{app:Commbound}
We now prove proposition \ref{prop:nncomm} in section \ref{sec:Fp}. Namely, for two consecutive boundary plaquettes $i$, $i+1$ and state $|\psi\rangle \in  {\cal V}^c$ such that the plaquettes $i$ and $i+1$ have the same spin $\tau^z_i = \tau^z_{i+1}$,  $F_i F_{i+1} |\psi\rangle = b F_{i+1} F_i |\psi\rangle$. The factor $b =  1$ if the plaquettes $i-1$ and $i+2$ have the same spin $\tau^z_{i-1} = \tau^z_{i+2}$. If they have opposite spin, then  $b = 1/\sqrt{2}$ if $\tau^z_{i-1} \neq \tau^z_{i} = \tau^z_{i+1} = \tau^z_{i+2}$, and $b = \sqrt{2}$ if $\tau^z_{i+2} \neq \tau^z_{i-1} = \tau^z_{i} = \tau^z_{i+1}$. 

\begin{figure}[h!]
\centering
\includegraphics[width=\textwidth]{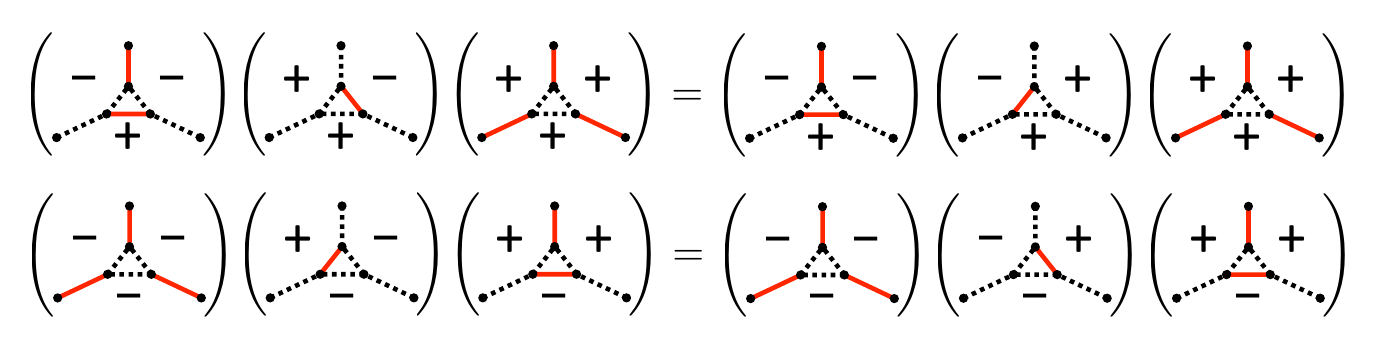}
\caption{Minimal identities of projectors needed to prove the identity, Eq.~(\ref{Piboundid}), related to the commutation relations of boundary plaquette flips. Left and right plaquettes are boundary plaquettes $i$ and $i+1$ respectively. Same notation as in figure \ref{fig:rules}.
}
\label{fig:rulesbound}
\end{figure}

{\it Proof.} We proceed in the same way as in section \ref{app:Commbulk}. The only difference is that the region $R_{i,i+1}$ is now as shown on the upper-left part of Fig.~\ref{fig:Rbound}. We need to prove the analogue of Eq.~(\ref{NPi3}), 
\beq N_{i,c_{i+1}}N_{i+1,c}  \Pi^{in}_{i,c_{i+1}} \Pi^{in}_{i+1,c} \Pi^{in}_c = b N_{i+1,c_{i}}N_{i,c}  \Pi^{in}_{i+1,c_i}\Pi^{in}_{i,c} \Pi^{in}_c \label{NPi3bound}\eeq
We will begin by showing  
\beq \Pi^{in}_{i,c_{i+1}} \Pi^{in}_{i+1,c} \Pi^{in}_c = \Pi^{in}_{i+1,c_i}\Pi^{in}_{i,c} \Pi^{in}_c \label{Piboundid} \eeq
i.e. the proportionality factor $b$ comes entirely from the normalization factors $N$,
\beq b = \frac{N_{i,c_{i+1}}N_{i+1,c}}{N_{i+1,c_{i}}N_{i,c}} \label{bN}\eeq
To prove (\ref{Piboundid}), there are two cases to consider, see Fig.~\ref{fig:rulesbound} - these identities are just hermitian conjugates of each other and can be proved by brute force.

It remains to compute the constant $b$ in Eq.~(\ref{bN}). We recall that for a boundary plaquette $j$, the normalization factor $N_{j,\tilde{c}} = 2^{L_{j, \tilde{c}}/4 - 1/2}$ if $\tau^z_{j-1} = \tau^z_{j+1}$, and $N_{j,\tilde{c}} = 2^{L_{j,\tilde{c}}/4-1/4}$ if $\tau^z_{j-1} \neq \tau^z_{j+1}$ in the spin configuration $\tilde{c}$, see Eqs.~(\ref{Xbound1}), (\ref{Xbound2}), (\ref{Xbound3}). Here, $L_{j, \tilde{c}}$ is the number of Majoranas in the segment ${\cal D}_{\tilde{c}} + {\cal D}_{\tilde{c}_j}$.  A key fact is that the sum of segment lengths  entering $F_i F_{i+1}$ and $F_{i+1} F_i$ is the same. Indeed, for the first row in Fig.~\ref{fig:rulesbound}, $L_{i, c_{i+1}} = L_{i, c} +1$, $L_{i+1, c_i} = L_{i+1, c}+1$, while for the second row, $L_{i, c_{i+1}} = L_{i, c} -1$, $L_{i+1, c_i} = L_{i+1, c}-1$. So, in either case, 
\beq L_{i,c_{i+1}} + L_{i+1,c} = L_{i+1,c_i} + L_{i,c} \label{Lsum}\eeq
 With this in mind, consider first the case when $\tau^z_{i-1} = \tau^z_{i} =\tau^z_{i+1} = \tau^z_{i+2}$. We then have, $N_{i+1,c} = 2^{L_{i+1, c}/4-1/2}$, $N_{i,c_{i+1}} = 2^{L_{i,c_{i+1}}/4-1/4}$, $N_{i,c} =  2^{L_{i, c}/4-1/2}$, $N_{i+1,c_i} =  2^{L_{i+1, c_i}/4-1/4}$, so using Eq.~(\ref{Lsum}), we obtain $b = 1$. 
 On the other hand, if $\tau^z_{i-1} \neq \tau^z_i = \tau^z_{i+1} = \tau^z_{i+2}$ then $N_{i+1,c} = 2^{L_{i+1, c}/4-1/2}$, $N_{i,c_{i+1}} = 2^{L_{i,c_{i+1}}/4-1/2}$, $N_{i,c} =  2^{L_{i, c}/4-1/4}$, $N_{i+1,c_i} =  2^{L_{i+1, c_i}/4-1/4}$, and $b = 1/\sqrt{2}$. The other two cases in proposition  \ref{prop:nncomm}  of section \ref{sec:Fp} can be analyzed in a similar manner.

\section{Boundary plaquette flip operators in bosonized notation}
\label{app:Fbos}
Here we illustrate the strategy to compute the matrix elements of plaquette flip operators $F_i$ in the bosonized notation. A useful observation is that we can choose a convenient base-point $B$. Indeed, we can use (\ref{Bchange}) to shift the base-point from $B$ to $B'$, act with $F_i$ and then shift the base-point back to $B$; since $F_i$ does not change the fermion parity the phase factors accumulated in the process cancel. 

We focus on the last ($c_8$) line in Eq.~(\ref{H3}).  To be specific, let's consider the case of NS spin structure and  odd fermion parity - other cases can be analyzed in a similar manner. We are interested in $F_i |\psi\rangle$ with
\beq |\psi \rangle =  \frac{1}{\sqrt{2}} (|a; (B-1, B), S_{\Phi} = -1\rangle -  |\bar{a}; (B-1, B); S_{\Phi} = -1\rangle) \eeq
Suppose $i$ belongs to a ``$-$" domain in the string $a$ stretching from $i_1$ to $j_1$. We choose the base-point $B = i_1$. Let the other ``$-$" domains be $(i_l, j_l)$ with $l =2 \ldots N_d$ arranged consecutively clockwise around the circle (more precisely, the double cover of the circle). We have,
\beq |\psi\rangle = (-1)^{a_{i_1-1}} \gamma_{i_1-1,i_1}^{a_{i_1-1} +a_{j_1+1}} F_{(i_1, j_1)} |\phi\rangle\eeq 
with 
\beq |\phi\rangle = \prod_{l=2}^{N_d} (s_{(i_1, i_l-1)} \gamma_{i_l-1,i_l})^{a_{i_l-1} +a_{j_l+1}} \prod_{l = 2}^{N_d} F_{(i_l, j_l)} |+\rangle \label{phi}\eeq
and terms in the first product in Eq.~(\ref{phi}) arranged with the smaller $l$'s to the left. Writing 
$F_{(i_1,j_1)} = F_{(i+2,j_1)} F_{i+1} F_i F_{(i_1,i-1)}$ and using, $[F_i, \gamma_{i_1-1,i_1}] = 0$, $[F_i, F_{(i+2,j_1)}] = 0$, 
\beq F_i |\psi\rangle = (-1)^{a_{i_1-1}} \gamma_{i_1-1,i_1}^{a_{i_1-1} +a_{j_1+1}} F_{(i+2,j_1)} F_i F_{i+1} F_i F_{(i_1,i-1)} |\phi\rangle \label{Fipsi1} \eeq
Now, from Proposition \ref{prop:nncomm} of section \ref{sec:Fp} we have $F_{i+1}  F_i F_{(i_1,i-1)} |\phi\rangle = \sqrt{2}   F_i F_{i+1}   F_{(i_1,i-1)} |\phi\rangle$, so
\beq F_i F_{i+1} F_i F_{(i_1,i-1)} |\phi\rangle = \sqrt{2} F^2_i F_{i+1}   F_{(i_1,i-1)} |\phi\rangle = \frac{1}{\sqrt{2}} (1+ i s_{i} \gamma_{i-1,i} \gamma_{i,i+1})F_{i+1}   F_{(i_1,i-1)} |\phi\rangle \eeq
where we used Proposition \ref{prop:square} of section \ref{sec:Fp} in the last step. Since $i s_{(i_1, i-1)} \gamma_{i_1-1,i_1} \gamma_{i-1,i} \sim 1$ on the state $F_{i+1}   F_{(i_1,i-1)} |\phi\rangle$,
\beq F_i F_{i+1} F_i F_{(i_1,i-1)} |\phi\rangle = \frac{1}{\sqrt{2}} (1+  s_{(i_1,i)} \gamma_{i_1-1,i_1} \gamma_{i,i+1})F_{i+1}   F_{(i_1,i-1)} |\phi\rangle\eeq
so returning to Eq.~(\ref{Fipsi1}),
\beq F_i |\psi\rangle = \frac{(-1)^{a_{i_1-1}}}{\sqrt{2}} \left( \gamma_{i_1-1,i_1}^{a_{i_1-1} +a_{j_1+1}} +  \gamma_{i_1-1,i_1}^{a_{i_1-1} +a_{j_1+1}+1} (s_{(i_1,i)} \gamma_{i,i+1})\right) F_{(i+1,j_1)}   F_{(i_1,i-1)} |\phi\rangle \label{Fipsi2}\eeq
Thus, the $(i_1,j_1)$ ``$-$" domain in $|\psi\rangle$ gets split into two ``$-$" domains $(i_1, i-1)$ and $(i+1,j_1)$ in $F_i |\psi\rangle$. Further, each of the two terms on the RHS of (\ref{Fipsi2}) (coming from the two terms in brackets) is in the canonical form (\ref{nstring}) (with basepoint $B = i_1$). Solving for $\{a'_j\}$ corresponding to the two terms in (\ref{Fipsi2}), we find that the first term has:  $a'_i = a_{j_1+1}$, $a'_{i+N} = \overline{a'_i}$ and all other $a'_j  = a_j$, while the second term has: $a'_i = 1-a_{j_1+1}$, $a'_{i+N} = \overline{a'_i}$ and all other $a'_j  =a_j$. Equivalently, we can say that one of the two terms has $a'_i  = 0$  and $a'_{i+N} = 1$, while the other term has $a'_i = 1$ and $a'_{i+N} = 0$ (and all other $a'_j  =a_j$); calling these two $a'$'s: $a^{i \to 0, i+N \to 1}$ and $a^{i \to 1, i+N \to 0}$ and noting that both of them have $a'_{i_1-1} = a_{i_1-1}$,
\beq F_i |\psi\rangle = \frac{1}{2} (|a^{i \to 0, i+N \to 1}\rangle - |\overline{a^{i \to 0, i+N \to 1}}\rangle) + \frac{1}{2} (|a^{i \to 1, i+N \to 0}\rangle - |\overline{a^{i \to 1, i+N \to 0}}\rangle) \label{Fipsi3}\eeq
where we have suppressed the labels $(B-1,B); S_\Phi = -1$, in each ket. We see that (\ref{Fipsi3}) exactly agrees with the $c_8$ term in Eq.~(\ref{H3}) with $c_8 =1$.

\section{Boundary Majorana operators in bosonized notation}
\label{app:Gammabos}
We illustrate the strategy for deriving the bosonized form (\ref{Gammapm}) of the Majorana operator $\Gamma^+_{i,i+1}$ (\ref{Gammap}). We  note that we can again choose a convenient base-point $B$: indeed, the dependence of states on $B$, Eq.~(\ref{Bchange}), cancels with the dependence of Eq.~(\ref{Gammap}) on $B$. 

Let us consider the case of NS spin structure and state $|\psi\rangle$ with $(-1)^{\cal F} = 1$. Other cases can be analyzed in a similar manner. We have
\beq |\psi\rangle = \frac{1}{\sqrt{2}} (|a; (B-1,B); S_{\Phi} =1\rangle +|\bar{a}; (B-1,B); S_{\Phi} =1\rangle)\eeq
$\Gamma^+_{i, i+1} |\psi\rangle$ does not vanish only if sites $i$ and $i+1$ have opposite spins. Let's first consider the case when site $i$ has $``+"$ spin and site $i+1$ has $``-"$ spin. Let us choose the base-point $B = i+1$. Going clockwise around the circle starting with the plaquette $B = i+1$, let the positions of  consecutive $``-"$ domains be $(i_l, j_l)$, $l = 1 \ldots N_d$, with $i_1 = i+1$. We have
\beq |\psi \rangle =  \prod_{l=1}^{N_d} (s_{(i_1, i_l-1)} \gamma_{i_l-1,i_l})^{a_{i_l-1} +a_{j_l+1}} \prod_{l = 1}^{N_d} F_{(i_l, j_l)} |+\rangle \label{psiGamma}\eeq
Then
\beq \Gamma^+_{i,i+1}|\psi\rangle = \gamma_{i,i+1}^{a_{i}+a_{j_1+1}+1} \prod_{l=2}^{N_d} (s_{(i_1, i_l-1)} \gamma_{i_l-1,i_l})^{a_{i_l-1} +a_{j_l+1} } \prod_{l = 1}^{N_d} F_{(i_l, j_l)} |+\rangle\eeq
The above expression for  $\Gamma^+_{i,i+1}|\psi\rangle$ is in the canonical form (\ref{nstring}). Let's work out the string $\{a'_j\}$ corresponding to it. Since $\Gamma^+_{i,i+1}|\psi\rangle$ has $(-1)^{\cal F} = -1$, $\{a'\}$ lives on the double-cover of the circle and has $a'_{j+N} = \overline{a'_j}$. We can  think of $a$ as also leaving on the double-cover of the circle with $a_{j+N} = a_j$. Then,  $a'_{i+k} = \overline{a_{i+k}}$, $k = 1\ldots N$, and $a'_{j} = a_j$ otherwise. In particular, $a'_i = a_i$. Therefore, from Eq.~(\ref{adefNSodd}), 
\beq \Gamma^+_{i,i+1}|\psi\rangle = \frac{(-1)^{a'_i}}{\sqrt{2}} \left(|a'; (B-1, B); S_{\Phi} = -1\rangle - |\overline{a'}; (B-1, B); S_{\Phi} = -1\rangle\right)\eeq
 This exactly agrees with the action of the first term of Eq.~(\ref{Gammapm}) on $|\psi\rangle$. 
 
 Next, let's consider the case when site $i$ has $``-"$ spin and site $i+1$ has $``+"$ spin. Let $i$ be the rightmost plaquette of a $``-"$ domain $(i_1, j_1)$ with $j_1 = i$. Let the remaining domains be $(i_l, j_l)$, $l =2\ldots N_d$ arranged clockwise consecutively around the circle. Let's choose the basepoint $B = i_1$. $|\psi\rangle$ again has the form (\ref{psiGamma}). Then,
 \beq \Gamma^+_{i,i+1} |\psi\rangle = s_{(i_1,i)} \gamma_{i,i+1} \gamma_{i_1-1,i_1}^{a_{i_1-1} + a_{i+1}} |\phi\rangle \eeq
 with
 \beq |\phi\rangle = \prod_{l=2}^{N_d} (s_{(i_1, i_l-1)} \gamma_{i_l-1,i_l})^{a_{i_l-1} +a_{j_l+1} } \prod_{l = 1}^{N_d} F_{(i_l, j_l)} |+\rangle\eeq
 Now, $i s_{(i_1, i)} \gamma_{i_1-1, i_1} \gamma_{i,i+1}|\phi\rangle = |\phi\rangle$, so
 \beq  \Gamma^+_{i,i+1} |\psi\rangle = i (-1)^{a_{i_1 -1} + a_{i+1} +1} \gamma_{i_1-1,i_1}^{a_{i_1-1} + a_{i+1} + 1} |\phi\rangle \eeq
 Again, $\Gamma^+_{i,i+1}|\psi\rangle$ is in the canonical form (\ref{nstring}). The corresponding string $\{a'_j\}$ on the double-cover of the circle again has $a'_{i+k} = \bar{a}_{i+k}$ for $k  =1 \ldots N$ and $a'_j = a_j$ otherwise. Noting that $a'_{i_1-1} = a_{i_1-1}$ and $a'_{i+1} = 1- a_{i+1}$, we have
 \beq \Gamma^+_{i,i+1} |\psi\rangle = i (-1)^{a'_{i+1}} \left(|a'; (B, B-1); S_{\Phi} = -1\rangle - |\overline{a'}; (B, B-1); S_{\Phi} = -1\rangle\right)\eeq
which exactly agrees with the action of the second term of Eq.~(\ref{Gammapm}) on $|\psi\rangle$.

\bibliography{Maj}

\end{document}